\documentclass[aps,prl,superscriptaddress]{revtex4}

\usepackage[dvips]{graphicx}
\usepackage{array}

\usepackage{amsmath}
\usepackage{amssymb}

\usepackage{amsmath}
\usepackage{amssymb}

\def\vone{{\bf l}} 
\def\bdm{\begin{displaymath}} \def\edm{\end{displaymath}}
\def\nn{\nonumber} \def\bc{\begin{center}} \def\ec{\end{center}}
\def\be{\begin{equation}} \def\ee{\end{equation}}

  \def\vA{{\bf A}} 
   
\def\vd{{\bf d}}  \def\vD{{\bf D}} \def\ve{{\bf e}}
 \def\vE{{\bf E}}  
\def\vg{{\bf g}}   
  \def\vj{{\bf j}} 
\def\vk{{\bf k}}   
   \def\vM{{\bf M}}
\def\vn{{\bf n}}   \def\vone{{\bf 1}}
\def\vp{{\bf p}}   \def\vq{{\bf q}}
 \def\vr{{\bf r}}  
   
\def\vU{{\bf U}} \def\vv{{\bf v}}  
 \def\vx{{\bf x}}  \def\vy{{\bf y}}
 \def\vz{{\bf z}} \def\v0{{\bf 0}} 
  \def\kB{k_{\rm B}}
\def\kF{k_{\rm F}} \def\EF{E_{\rm F}} \def\NF{N_{\rm F}} 
  \def\vna{{\bf\nabla}} 
\def\vgamma{{\mbox{\boldmath$\gamma$}}} \def\vsigma{{\mbox{\boldmath$\sigma$}}}
\def\vtau{{\mbox{\boldmath$\tau$}}}
\def\vDelta{{\mbox{\boldmath$\Delta$}}}

\def\vna{{\bf\nabla}}
\def\vxi{{\mbox{\boldmath$\xi$}}}

\def\vDelta{{\mbox{\boldmath$\Delta$}}}

\def\kB{k_{\rm B}} \def\kF{k_{\rm F}} \def\EF{E_{\rm F}} \def\NF{N_{\rm F}}

\makeindex             

\begin{document}

\title{Kinetic theory for response and transport in non--centrosymmetric superconductors}

\author{Ludwig Klam}
\email[]{L.Klam@fkf.mpg.de}
\affiliation{Max-Planck-Institut f\"{u}r Festk\"{o}rperforschung, Heisenbergstrasse 1, D-70569 Stuttgart}

\author{Dirk Manske}
\affiliation{Max-Planck-Institut f\"{u}r Festk\"{o}rperforschung, Heisenbergstrasse 1, D-70569 Stuttgart}

\author{Dietrich Einzel}
\affiliation{Walther-Meissner-Institut, Bayerische Akademie der Wissenschaften, D-85748 Garching}

\begin{abstract}
We formulate a kinetic theory for non--centrosymmetric superconductors at low temperatures in the clean limit. The transport equations are solved quite generally in spin-- and particle--hole (Nambu) space by performing first a transformation into the band basis and second a Bogoliubov transformation to the quasiparticle--quasihole phase space. Our result is a particle--hole--symmetric, gauge--invariant and charge conserving description, which is valid in the whole quasiclassical regime ($|\vq|\ll k_F$ and $\hbar\omega\ll \EF$). We calculate the current response, the specific heat capacity, and the Raman response function.
For the Raman case, we investigate within this framework the polarization--dependence of the electronic (pair--breaking) Raman response for the recently discovered non--centrosymmetric superconductors at zero temperature. Possible applications include the systems CePt$_3$Si and Li$_2$Pd$_x$Pt$_{3-x}$B, which reflect the two important classes of the involved spin--orbit coupling. We provide analytical expressions for the Raman vertices for these two classes and calculate the polarization--dependence of the electronic spectra. We predict a two--peak structure and different power laws with respect to the unknown relative magnitude of the singlet and triplet contributions to the superconducting order parameter, revealing a large variety of characteristic fingerprints of the underlying condensate.
\end{abstract}

\maketitle



\section{Introduction}
\label{sec:Intro}
In a large class of conventional and in particular unconventional superconductors a classification
of the order parameter with respect to spin singlet/even parity and spin triplet/odd parity is possible,
using the Pauli exclusion principle. A necessary prerequisite for such a classification is, however,
the existence of an inversion center. Something of a stir has been caused by the discovery of the bulk
superconductor CePt$_3$Si without inversion symmetry \cite{Bauer:2004:01}, which initiated extensive theoretical \cite{Frigeri:2004:02,Samokhin:2008:01} and experimental studies \cite{Bauer:2005:02,Fak:2008:02}. In such systems the existence of an antisymmetric potential gradient causes a parity--breaking antisymmetric spin--orbit coupling (ASOC), that gives rise to the possibility of having admixtures of spin--singlet and spin--triplet pairing states. Such parity--violated, non--centrosymmetric superconductors (NCS) are the topic of this chapter, which is dedicated particularly to a theoretical study of response and transport properties at low temperatures.
We will use the framework of a kinetic theory described by a set of generalized Boltzmann equations, successfully used before in \cite{Einzel:2008:02}, to derive various response and transport functions such as the normal and superfluid density, the specific heat capacity (i. e. normal fraction and condensate properties, that are native close to the long wavelength, stationary limit) and in particular the electronic Raman response in NCS (which involves frequencies $\hbar\omega$ comparable to the energy gap 
$\Delta_\vk$ of the superconductor).

A few general remarks about the connection between response and transport phenomena are appropriate at this stage.
Traditionally, the notion of transport implies that the theoretical description takes into account the effects of
quasiparticle scattering processes, represented, say,  by a scattering rate $\Gamma$. Therefore, we would like to 
demonstrate with a simple example, how response and transport are intimately connected: consider the density 
response of normal metal electrons to the presence of the two electromagnetic potentials 
$\Phi^{\rm ext}$ and $\vA^{\rm ext}$, which generate the gauge--invariant form of the electric field $\vE=-\vna\Phi^{\rm ext}-\partial\vA/c\partial t$. In Fourier space ($\vna\rightarrow i\vq$,
$\partial/\partial t\rightarrow -i\omega$) one may write for the response of the charge density:
\begin{eqnarray*}
\delta n_e=e^2i\vq\cdot\vM_0(\vq,\omega)\cdot\vE
\end{eqnarray*}
with $\vM_0$ the Lindhard tensor and $\vq\cdot\vM_0\cdot\vq\equiv M_0$ the Lindhard function, appropriately 
renormalized by collision effects \cite{Mermin:1970:01}:
\begin{eqnarray*}
M_0(\vq,\omega)=\frac{{\cal L}_0(\vq,\omega+i\Gamma)}{1-\frac{i\Gamma}{\omega+i\Gamma}\left[1-\frac{{\cal L}_0(\vq,\omega+i\Gamma)}{{\cal L}_0(\vq,0)}\right]}
\end{eqnarray*}
Here ${\cal L}_0(q,\omega)$ denotes the unrenormalized Lindhard function in the collisionless limit
$\Gamma\rightarrow 0$:  
\begin{eqnarray*}
{\cal L}_0(\vq,\omega)=\frac{1}{V}\sum_{\vp\sigma}
\frac{n^0_{\vp+\vq/2}-n^0_{\vp-\vq/2}}{\epsilon_{\vp+\vq/2}-\epsilon_{\vp-\vq/2}-\hbar\omega}
\; .
\end{eqnarray*}
In this definition of the Lindhard function, $n_\vk^0$ denotes the equilibrium Fermi--Dirac distribution function and $\epsilon_\vk=\xi_\vk+\mu$ represents the band structure. Now the aspect of transport comes into play by the observation that $M_0(\vq,\omega+i\Gamma)$ may be expressed through the full dynamic conductivity tensor $\vsigma(\vq,\omega)
=e^2(\partial n/\partial\mu)\vD(\vq,\omega)$ of the electron system as follows:
\begin{eqnarray*}
M_0(\vq,\omega)\equiv\frac{\vq\cdot\vsigma(\vq,\omega)\cdot\vq}{i\omega-\vq\cdot\vD(\vq,\omega)\cdot\vq/(1-i\omega\tau)}
\end{eqnarray*}
with $\vq\cdot\vsigma\cdot\vq\stackrel{\Gamma\to0}{\equiv} i\omega e^2{\cal L}_0(\vq,\omega)$ and
with the so--called diffusion pole in the denominator of $M_0(\vq,\omega)$ reflecting the 
charge conservation law. This expression for the Lindhard response function $M_0$ clearly demonstrates the 
connection between response (represented by $M_0$ itself) and transport (represented by the conductivity $\vsigma$), which 
can be evaluated both in the clean limit $\Gamma\rightarrow 0$ and in the presence of collisions $\Gamma\not=0$.
In this sense, the notions of response and transport are closely connected and therefore equitable. In this whole 
chapter we shall limit or considerations to the collisionless case.

An important example for a response phenomenon involving finite frequencies is the electronic Raman effect. Of particular interest is the so--called pair--breaking Raman effect, in which an incoming photon breaks a Cooper pair of energy $2\Delta_{\vk}$ on the Fermi surface, and a scattered photon leaves the sample with a frequency reduced by $2\Delta_{\vk}/\hbar$, has turned out to be a very effective tool to study unconventional superconductors with gap nodes. This is because various choices of the photon polarization with respect to the location of the nodes on the Fermi surface allow one to draw conclusions about the node topology and hence the pairing
symmetry. An example for the success of such an analysis is the important work by Devereaux {\it et al.} \cite{Devereaux:1994:01} in which the $d_{x^2-y^2}$--symmetry of the order parameter in cuprate superconductors could be traced back to the frequency--dependence of the electronic Raman spectra, that directly measured the pair--breaking effect. Various theoretical studies of NCS have revealed a very rich and complex node structure in parity--mixed order parameters, which can give rise to qualitatively very different shapes, i. e. frequency dependencies, of the Raman intensities, ranging from threshold-- and cusp-- to singularity--like behavior. Therefore the study of the polarization dependence of Raman spectra enables one to draw conclusions about the internal structure of the parity--mixed gap parameter in a given NCS.

This chapter is organized as follows:
In section~\ref{subsec:Intro:ASOC} we introduce our model for the ASOC, the two order parameters on the spin--orbit splitted bands and the pairing interaction.
Then, in section~\ref{sec:Transp} we derive the kinetic transport equations for NCS at low temperatures in the clean limit and transform these equations into the more convenient band--basis.
In section~\ref{sec:solution}, the transport equations are solved quite generally in band-- and particle--hole (Nambu) space by first performing a Bogoliubov transformation to the quasiparticle--quasihole phase space and second performing the inverse Bogoliubov transformation to recover the original distribution functions.
We demonstrate gauge invariance of our theory in section~\ref{subsec:Results:gaugemode} by taking the fluctuations of the order parameter into account.
Within this framework, we calculate the normal and superfluid density in section~\ref{subsec:Results:superfl_density} and the specific heat capacity in section~\ref{subsec:Results:heatcapacity}.
In section~\ref{subsec:Results:Raman}, our particular interest is focussed on the electronic Raman response.
We investigate the polarization--dependence of the pair--breaking Raman response at zero temperature for two important classes of the involved spin--orbit coupling.
Finally, in section~\ref{sec:Conclusion} we summarize our results and draw our conclusions.

\section{Antisymmetric spin--orbit coupling}
\label{subsec:Intro:ASOC}

\begin{figure}[t] 
\begin{minipage}[c]{0.45\linewidth}
   \noindent \hspace{-1.0\linewidth} (a)\\
  \includegraphics[angle=0, width=0.9\linewidth]{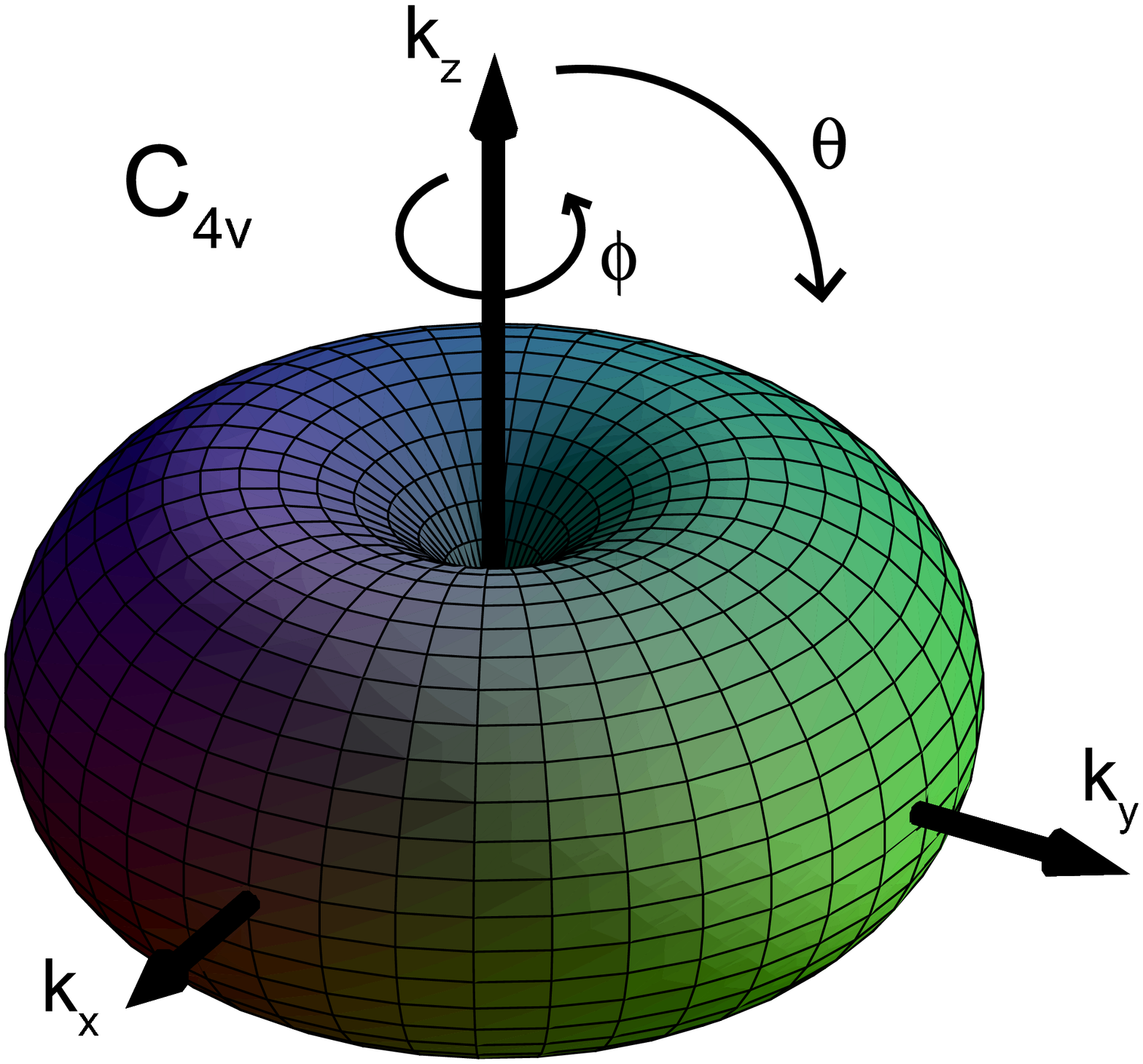}
\end{minipage}
\hspace{0.04\linewidth}
\begin{minipage}[c]{0.45\linewidth}
 \noindent \hspace{-1.0\linewidth} (b)\\
 \includegraphics[angle=0, width=0.9\linewidth]{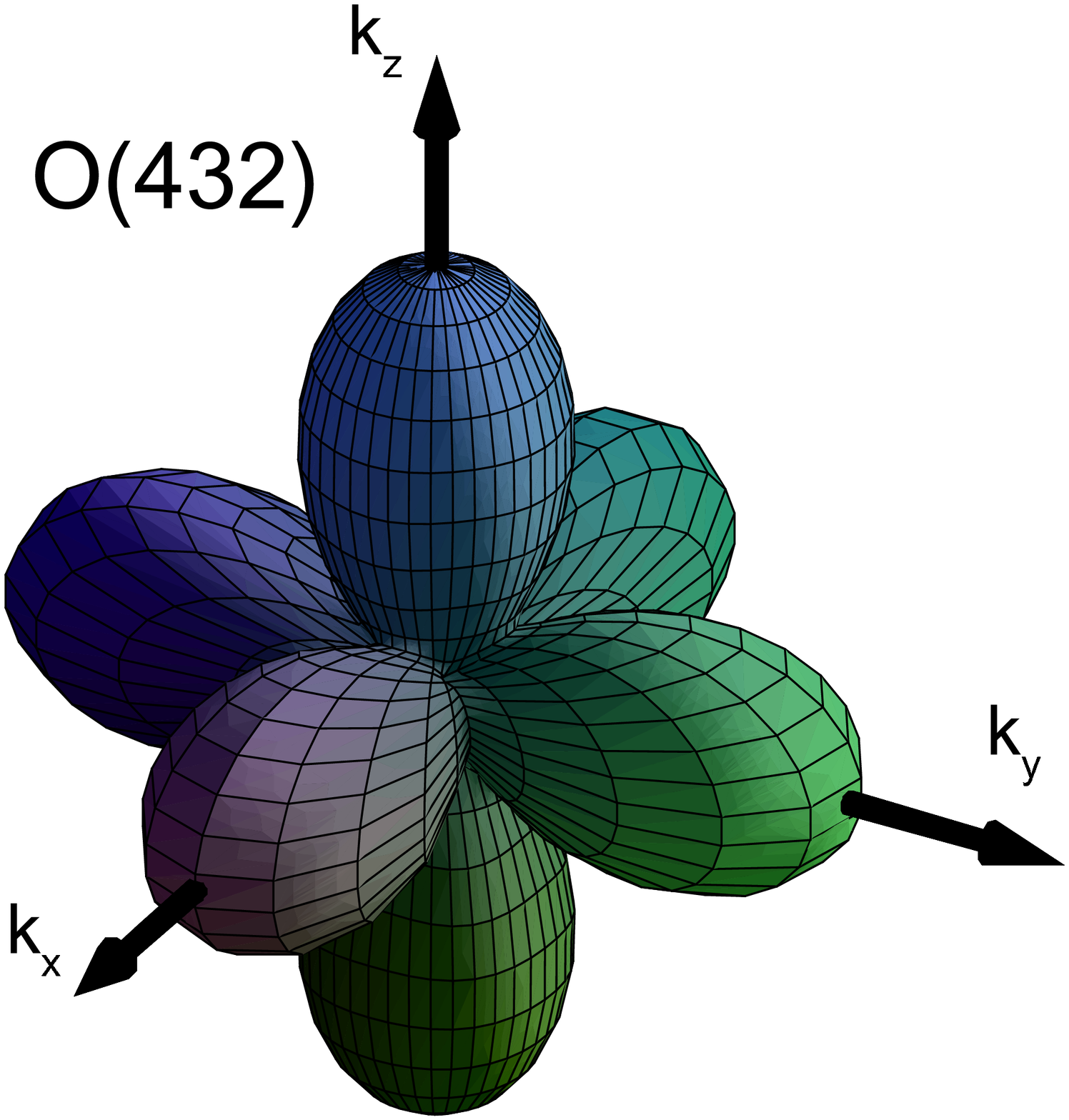}
\end{minipage}
\caption{The angular dependence of $|\vgamma_\vk|$ for the point groups C$_{4v}$ and $O$. Since $\vd_\vk || \vgamma_\vk$, these plots show also the magnitude of the gap function in the pure triplet case for both point groups.}
\label{fig:gk}
\end{figure}

We start from a model Hamiltonian for noninteracting electrons in a non--centro\-symmetric crystal \cite{Samokhin:2007:01}
\begin{eqnarray}
\hat H &=& \sum_{\vk\sigma\sigma^\prime} \hat c^\dagger_{\vk\sigma}\left[ \xi_\vk \delta_{\sigma\sigma^\prime} + \vgamma_\vk\cdot\vtau_{\sigma\sigma^\prime} \right] \hat c_{k\sigma^\prime} \; , \label{Eq_Hamilton}
\end{eqnarray}
where $\xi_\vk$ represents the bare band dispersion assuming time reversal symmetry ($\xi_{-\vk}=\xi_\vk$), $\sigma,\sigma^\prime=\uparrow,\downarrow$ label the spin state, and $\vtau$ are the Pauli matrices. The second term describes an antisymmetric spin--orbit coupling (ASOC) with a (vectorial) coupling constant $\vgamma_\vk$. The pseudovector function $\vgamma_\vk$ has the following symmetry properties: $\vgamma_{-\vk}=-\vgamma_\vk$ and $g\vgamma_{g^{-1}\vk}=\vgamma_\vk$.
Here $g$ denotes any symmetry operation of the point group $\mathcal{G}$ of the crystal under consideration.
In NCSs two important classes of ASOCs are realized, reflecting the underlying point group $\mathcal{G}$ of the crystal.
In particular, we shall be interested in the tetragonal point group $C_{4v}$ (applicable to the heavy Fermion compound CePt$_3$Si with T$_c$=0.75 K \cite{Bauer:2004:01} for example) and the cubic point group $O$ (applicable to the system Li$_2$Pd$_x$Pt$_{3-x}$B with T$_{c}$=2.2--2.8 K for x=0 and T$_{c}$=7.2--8 K for x=3 \cite{Badica:2004:01}).
For $\mathcal{G}=C_{4v}$ the ASOC reads \cite{Frigeri:2004:01,Samokhin:2007:01}
\begin{equation}
  \vgamma_\vk = \mathrm{g}_\bot ( \hat\vk\times\hat\ve_z ) + \mathrm{g}_\Vert \hat k_x\hat k_y\hat k_z ( \hat k^2_x - \hat k^2_y )\hat \ve_z \label{Eq_gk_tetragonal} \; .
\end{equation}
In the purely two--dimensional case ($\mathrm{g}_\Vert=0$) one recovers, what is known as the Rashba interaction \cite{Dresselhaus:1955:01,Edelstein:1989:01,Gorkov:2001:01}. We will choose for simplicity $\mathrm{g}_\Vert=0$ for our Raman results.
For the cubic point group $\mathcal{G}=O$, $\vgamma_\vk$ reads \cite{Yuan:2006:01}
\begin{equation}
  \vgamma_\vk = \mathrm{g}_1 \hat \vk - \mathrm{g}_3 \left[ \hat k_x (\hat k_y^2 + \hat k_z^2)\hat \ve_x + \hat k_y (\hat k_z^2 + \hat k_x^2)\hat \ve_y + \hat k_z (\hat k_x^2 + \hat k_y^2)\hat \ve_z \right] \; , \label{Eq_gk_cubic}
\end{equation}
where the ratio $\mathrm{g}_3/\mathrm{g}_1\simeq3/2$ is estimated by Ref.~\cite{Yuan:2006:01}.
Because of the larger prefactor $\mathrm{g}_3>\mathrm{g}_1$ we will keep the higher order term for our further considerations.
Thus, in terms of spherical angles, $\hat\vk= (\cos\phi\sin\theta$, $\sin\phi\sin\theta$, $\cos\theta)$, the absolute value of the $\vgamma_\vk$--vectors for both point groups, illustrated in Fig.~\ref{fig:gk}, reads
\begin{align}
  |\vgamma_\vk| &= \sin\theta && \mbox{for\ \ $C_{4v}$}\\
  |\vgamma_\vk| &= \sqrt{1- \frac{15}{16}\sin^2 2\theta - \frac{3}{16} \sin^4 \theta \sin^2 2\phi \left( 9\sin^2 \theta - 4 \right)} && \mbox{for\ \ $O$}
\end{align}

By diagonalizing the Hamiltonian [Eq.~(\ref{Eq_Hamilton})], one finds the eigenvalues $\xi_{\lambda}(\vk)=\xi_\vk + \lambda|\vgamma_\vk|$, which physically corresponds to the lifting of the Kramers degeneracy between the two spin states at a given $\vk$ in the presence of ASOC. The basis in which the band is diagonal can be referred to as the {\it band basis} where the Fermi surface defined by $\xi_{\pm}(\vk)=0$ is splitted into two pieces labeled $\pm$.
Sigrist and co--workers have shown that the presence of the ASOC generally allows for an admixture of a spin--triplet component to the otherwise spin--singlet pairing gap \cite{Frigeri:2004:02}. This implies that we may write down the following ansatz for the energy gap matrix in spin space:
\begin{equation}
  \vDelta_{\sigma\sigma^\prime}(\vk) = \{[\psi_\vk (T) \vone + \vd_\vk (T)\cdot \vtau ] i \vtau^y\}_{\sigma\sigma^\prime}\; , \label{eq:ASOC:gap:spin}
\end{equation}
where $\psi_\vk(T)$ and $\vd_\vk(T)$ reflect the singlet and triplet part of the pair potential, respectively.
In the band basis we find immediately
\begin{equation}
  \Delta_{\pm}(\vk) = \psi_\vk (T) \pm |\vd_\vk (T)|\; . \label{eq:ASOC:gap:bandd}
\end{equation}
It has been demonstrated that a large ASOC compared to $\kB T_c$ is not destructive for triplet pairing if one assumes $\vd_\vk \Vert \vgamma_\vk$ \cite{Frigeri:2004:02,Samokhin:2008:01}:
\begin{equation}
  \vd_\vk (T) = d(T) \hat \vgamma_\vk \; ,
\end{equation}
whereas the temperature--dependent magnitudes $\psi(T)$ and $d(T)$ of the spin--singlet and triplet energy gaps are solutions of coupled self--consistency equations and $\hat \vgamma_\vk$ is defined by
\begin{equation}
  \hat \vgamma_\vk = \frac{\vgamma_\vk}{\sqrt{\langle|\vgamma_\vk|^2\rangle_{FS}}}\; .
\end{equation}
Thus the energy gap of Eq.~(\ref{eq:ASOC:gap:bandd}) can be written as:
\begin{equation}
  \Delta_{\pm}(\vk) = \psi(T) \pm d(T)|\hat\vgamma_\vk| \; .
\end{equation}
For the $T=0$ Raman response in section~\ref{subsec:Results:Raman} we will use the following ansatz for the gap function on both bands ($+$~and~$-$) \cite{Frigeri:2006:01}:
\begin{equation}
  \Delta_{\pm}(\vk) = \psi \pm d|\vgamma_\vk| = \psi \left( 1 \pm p|\vgamma_\vk| \right) \equiv \Delta_\pm \; , \label{Eq_gap}
\end{equation}
where the parameter $p=d/\psi$ represents the unknown triplet--singlet ratio.
Accordingly, the Bogoliubov--quasiparticle dispersion is given by $E^2_\pm(\vk)=\xi^2_\pm(\vk)+\Delta^2_\pm(\vk)$.
If we assume no $\vq$--dependence of the order parameter, $\Delta_\lambda(\vk)$ [and also $E_\lambda(\vk)$] is of even parity i.e. $\Delta_\lambda(-\vk)=\Delta_\lambda(\vk)$. It is quite remarkable that although the spin representation of the order parameter $\Delta_{\sigma\sigma^\prime}(\vk)$ has no well--defined parity w.r.t. $\vk\rightarrow -\vk$, as easily seen in Eq.~(\ref{eq:ASOC:gap:spin}), the energy gap in band representation has.
Note that for Li$_2$Pd$_x$Pt$_{3-x}$B the parameter $p$ seems to be directly related to the substitution of platinum by palladium, since the larger spin--orbit coupling of the heavier platinum is expected to enhance the triplet contribution \cite{Lee:2005:01}. This seems to be confirmed by penetration depth experiments \cite{Yuan:2005:01,Yuan:2006:01}.

The corresponding weak--coupling gap equation reads
\begin{equation}
    \Delta_\lambda(\vk,T) = -\sum\limits_{\vk^\prime,\mu} V^{\lambda\mu}_{\vk\vk^\prime} \Delta_\mu(\vk^\prime,T) \theta_\mu(\vk^\prime) \label{Eq_gap_eq}
\end{equation}
with
\begin{equation}
 \theta_{\lambda}(\vk) = \frac{1}{2E_{\lambda}(\vk)}\tanh\frac{E_{\lambda}(\vk)}{2\kB T}
\end{equation}
and its solution are extensively discussed in Ref.~\cite{Frigeri:2006:01}.
Here and in the following we choose a separable ansatz for the pairing--interaction (cf. Ref.~\cite{Frigeri:2006:01} with $e_m=0$, i.e. without Dzyaloshinskii--Moriya interaction):
\begin{equation}
    V^{\lambda\mu}_{\vk\vk^\prime} = \Gamma_s + \lambda\mu\Gamma_t|\hat\vgamma_\vk||\hat\vgamma_{\vk^\prime}| \; , \label{Eq_pairing_int}
\end{equation}
where $\Gamma_s$ and $\Gamma_t$ represent the singlet and triplet contribution, respectively.
Although an exact numerical solution of Eqs.~(\ref{Eq_gap_eq})--(\ref{Eq_pairing_int}) with a microscopic pairing interaction would be desirable, we restrict ourselves in this work to a phenomenological description which allows an analytical treatment of response and transport in NCS.

\section{Derivation of the transport equations}
\label{sec:Transp}
In this section, we study the linear response of the superconducting system to an effective external perturbation potential of the form
\begin{equation}
    \delta\xi^{\rm ext}_{\vk\sigma\sigma^\prime} = \left[ e\Phi(\vq,\omega) - \frac{e}{c}\vv_\vk\cdot\vA(\vq,\omega) \right] \delta_{\sigma\sigma^\prime} + \frac{e^2}{c^2}\vA_i^I(\vq,\omega)\frac{\partial^2\epsilon_\vk}{\hbar^2\partial\vk_i\partial\vk_j}\vA^S_j(\vq,\omega) \delta_{\sigma\sigma^\prime} \; . \label{eq:transp:xik}
\end{equation}
Here $\Phi$ and $\vA$ denote the electromagnetic scalar and vector potential.
Electronic Raman scattering is described in addition by the third term in Eq.~(\ref{eq:transp:xik}).
It describes a Raman process where an incoming photon with vector potential $\vA^I$, polarization $\hat\ve^I$ and frequency $\omega_I$ is scattered off an electronic excitation. The scattered photon with vector potential $\vA^S$, polarization $\hat\ve^S$ and frequency $\omega_S=\omega_I-\omega$ gives rise to a Raman signal (Stokes process) and creates an electronic excitation with momentum transfer $\vq$.
Further, the Raman vertex in the so--called effective--mass approximation reads
\begin{equation}
    \gamma^{(R)}_\vk = m \sum\limits_{i,j} \hat\ve^S_i \frac{\partial^2 \epsilon(\vk)}{\hbar^2\partial k_i \partial k_j} \hat\ve^I_j \; .
\end{equation}
In general, an external perturbation can be decomposed into a vertex function $a_\vk$ and a related potential $\delta\xi_a$:
\begin{equation}
    \delta\xi_\vk^{\rm ext} = \sum\limits_a a_\vk\delta\xi_a \; .
\end{equation}
A list of all relevant vertex functions and potentials that will be discussed in this chapter is given in Table~\ref{tab:pert}.
The charge density response to the electric field $\vE=-\vna\Phi^{\rm ext}-\partial\vA/c\partial t$ is characterized by a constant vertex $a_\vk=e$ (electron charge) and therefore of even parity (w.r.t. $\vk\rightarrow -\vk$), whereas the current response to the vector potential $\vA$ depends on the odd vertex--function $a_\vk=e\vv_\vk$ (electron velocity).
In case of the specific heat capacity $C_V(T)$, the role of the fictive potential is played by the temperature change $\delta T$, which couples to the energy variable $\xi_\vk$.
For the Raman response, this fictive potential depends essentially on the vector potential of the incoming and scattered light.
The response and transport functions will be obtained as moments of the momentum distribution functions with the corresponding vertex (see section~\ref{subsec:Results:superfl_density}--\ref{subsec:Results:Raman}).
%
%
\begin{table}
\caption{External perturbations can be decomposed into a vertex--function and a potential. The vertex--function is characteristic for each response--function and can be classified according to parity (w.r.t. $\vk\rightarrow -\vk$) and dimension.}
\label{tab:pert}
\begin{tabular}{p{3cm}p{2.5cm}p{1.5cm}p{2cm}p{2cm}p{2cm}}
\hline\noalign{\smallskip}
vertex & (fictive) potential & parity & dimension & response  \\
$a_\vk$ & $\delta\xi_a$ & \; & \; & \; \\
\hline\hline\noalign{\smallskip}
$e$  & $\Phi^{\rm ext}$ & even & scalar & charge and\\
$e\vv_\vk$ & $\vA^{\rm ext}$ & odd & vector & current\\
$\left( \frac{E_\lambda(\vk)}{T}- \frac{\partial E_\lambda(\vk)}{\partial T} \right)$ & $\delta T$ & even & scalar &  specific heat\\
\; & \; & \; & \; & \mbox{capacity}\\
$m(\vM^{-1}_\vk)_{i,j}$  & $r_0 A^I_i A^S_j$ & even & tensor & Raman\\
\noalign{\smallskip}\hline\noalign{\smallskip}
\end{tabular}
\end{table}

In addition to the external perturbation potentials, also molecular potentials can be taken into account within a mean--field approximation:
\begin{equation}
    \delta\xi_\vk = \delta\xi^{\rm ext}_\vk + \sum\limits_{\vp\sigma}\left( f^s_{\vk\vp} + V_\vq \right)\delta n_\vp = \sum\limits_a a_\vk\delta\xi_a + V_\vq\delta n_1 + \sum\limits_{\vp\sigma} f^s_{\vk\vp} \delta n_\vp \; .
\end{equation}
The short--range Fermi--liquid interaction $f^s_{\vk\vp}$ leads to a renormalization of the electron mass \cite{Pines:1966:01} and the long--ranged Coulomb interaction with $V_\vq=4\pi e^2/\vq^2$ is included self--consistently through the macroscopic density fluctuations $\delta n_1 = \sum_{\vp\sigma}\delta n_\vp$ with the non--equilibrium momentum distribution function $\delta n_\vp$.

The potentials $\delta\xi^{\rm ext}_\vk$ are assumed to vary in time and space $\propto exp(i\vq\cdot\vr - i\omega t)$. Then the response to the perturbation potentials can generally be described by a nonequilibrium momentum distribution function $\underline{n}_{\vp\vp^\prime}$, which is a matrix in Nambu, momentum and spin space with $\vp=\vk+\vq/2$ and $\vp^\prime= \vk-\vq/2$.
The evolution of the nonequilibrium matrix distribution function in time and space is governed by the matrix--kinetic (von Neumann) equation \cite{Betbeder:1969:01,Woelfle:1976:01}
\begin{equation}
    \hbar\omega \underline{n}_{\vp\vp^\prime} + \sum\limits_{\vp^{\prime\prime}} \left[ \underline{n}_{\vp\vp^{\prime\prime}}, \underline{\xi}_{\vp^{\prime\prime}\vp^\prime} \right]_- = 0 \label{eq:transp:neumann}
\end{equation}
in which the full quasiparticle energy $\underline{\xi}_{\vp\vp^\prime}$ plays the role of the Hamiltonian of the system.
This equation holds for $\hbar\omega\ll\EF$ and $|\vq|\ll k_{\rm F}$.
In general, a collision integral (see e.g. \cite{Woelfle:1976:01}) could be inserted on the right hand side of Eq.~(\ref{eq:transp:neumann}) that accounts for the relaxation of the system into local equilibrium through collisions. In the following we will assume the absence of collisions \footnote{An example for collision integrals in the Raman response can be found in Ref.~\cite{Einzel:2008:02}}.
After linearization according to
\begin{eqnarray}
    \underline{n}_{\vp^{\prime\prime}\vp^\prime}  &=& \underline{n}_\vk(\vq,\omega)  = \underline{n}^0_\vk \delta_{\vq,0} + \delta\underline{n}_\vk(\vq,\omega) \\
    \underline{\xi}_{\vp^{\prime\prime}\vp^\prime} &=& \underline{\xi}_\vk(\vq,\omega) = \underline{\xi}^0_\vk \delta_{\vq,0} + \delta\underline{\xi}_\vk(\vq,\omega) \;,
\end{eqnarray}
the matrix--kinetic equation assumes the following form in $\omega-\vq$-- and spin--space:
\begin{equation}
    \hbar\omega\delta\underline{n}_\vk + \delta\underline{n}_\vk\underline{\xi}^0_{\vk-} - \underline{\xi}^0_{\vk+}\delta\underline{n}_\vk = \delta\underline{\xi}_\vk \underline{n}^0_{\vk-} - \underline{n}^0_{\vk+}\delta\underline{\xi}_\vk \; . \label{eq:MKE:spin}
\end{equation}
Here, $\omega$ is the frequency and $\vk\pm=\vk\pm\vq/2$, with $\vq$ representing the wave number of the external perturbation.
The equilibrium distribution function $\underline{n}^0_\vk$ and quasiparticle energy $\underline{\xi}^0_\vk$ are matrices in Nambu and spin space:
\begin{eqnarray}
    \underline{n}^0_\vk &=& \left( \begin{array}{cc}
        \vn_\vk & \vg_\vk \\
        \vg^\dagger_\vk & \vone-\vn_{-\vk}
    \end{array} \right) \\
    \underline{\xi}^0_\vk &=& \left( \begin{array}{cc}
        \vxi_\vk+\vgamma_\vk\cdot\vtau & \vDelta_\vk \\
        \vDelta^\dagger_\vk & -\left[ \vxi_\vk+\vgamma_\vk\cdot\vtau \right]^T
    \end{array} \right) \; .
\end{eqnarray}
The momentum and frequency--dependent deviations from equilibrium are defined as
\begin{equation}
    \delta\underline{n}^0_\vk = \left( \begin{array}{cc}
        \delta\vn_\vk & \delta\vg_\vk \\
        \delta\vg^\dagger_\vk & -\delta\vn_{-\vk}
    \end{array} \right)
\end{equation}
and
\begin{equation}
    \delta\underline{\xi}^0_\vk = \left( \begin{array}{cc}
        \delta\vxi_\vk & \delta\vDelta_\vk \\
        \delta\vDelta^\dagger_\vk & -\delta\vxi_{-\vk}
    \end{array} \right)\; ,
\end{equation}
respectively.
In the spin basis, the matrix--kinetic equation [Eq.~(\ref{eq:MKE:spin})] represents a set of 16 equations, which can be reduced to a set of 8 equations by an unitary transformation into the band basis (also referred to as helicity basis).
This SU(2) rotation is given by \cite{Vorontsov:2008:01}
\begin{eqnarray}
 \underline{U}_\vk &=& \left( \begin{array}{cc} \vU_\vk & 0 \\ 0 & \vU^*_\vk \end{array} \right) \\
 \vU_\vk &=& \exp \left(-i\frac{\theta_\gamma}{2}\hat\vn_\gamma\cdot\vtau \right) = \cos \frac{\theta_\gamma}{2} - i\hat\vn_\gamma\cdot\vtau \sin \frac{\theta_\gamma}{2} \\
 \vn_\gamma &=& \frac{\vgamma_\vk\times\hat\vz}{|\vgamma_\vk\times\hat\vz|} \; ,
\end{eqnarray}
which corresponds to a rotation in spin space into the $\hat\vz$--direction about the polar angle $\theta_\gamma$ between $\vgamma_\vk$ and $\hat\vz$.
Then Eq.~(\ref{eq:MKE:spin}) may be written as
\begin{eqnarray}
 \hbar\omega\underline{U}^\dagger_{\vk+}\delta\underline{n}_\vk\underline{U}_{\vk-}
 &+& \underline{U}^\dagger_{\vk+}\delta\underline{n}_\vk \underline{U}_{\vk-}\underline{U}^\dagger_{\vk-} \underline{\xi}^0_{\vk-} \underline{U}_{\vk-}
   - \underline{U}^\dagger_{\vk+}\underline{\xi}^0_{\vk+} \underline{U}_{\vk+}\underline{U}^\dagger_{\vk+} \delta\underline{n}_\vk \underline{U}_{\vk-} \nn\\
 &=& \underline{U}^\dagger_{\vk+}\delta\underline{\xi}_\vk \underline{U}_{\vk-}\underline{U}^\dagger_{\vk-} \underline{n}^0_{\vk-} \underline{U}_{\vk-}
   - \underline{U}^\dagger_{\vk+}\underline{n}^0_{\vk+} \underline{U}_{\vk+}\underline{U}^\dagger_{\vk+} \delta\underline{\xi}_\vk \underline{U}_{\vk-}
\end{eqnarray}
or, more simply
\begin{equation}
 \hbar\omega\delta\underline{n}^b_\vk + \delta\underline{n}^b_\vk\underline{\xi}^b_{\vk-} - \underline{\xi}^b_{\vk+}\delta\underline{n}^b_\vk = \delta\underline{\xi}^b_\vk \underline{n}^b_{\vk-} - \underline{n}^b_{\vk+}\delta\underline{\xi}^b_\vk \; , \label{eq:MKE:band}
\end{equation}
where the equilibrium distribution function and energy shifts in the band basis are given by
\begin{equation}
 \underline{n}^b_\vk = \left( \begin{array}{cccc}
        \frac{1}{2}(1-\xi_{+}\theta_{+}) & 0 & 0 & -\Delta_{+}\theta_{+} \\
        0 & \frac{1}{2}(1-\xi_{-}\theta_{-}) & \Delta_{-}\theta_{-} & 0 \\
        0 & \Delta^*_{-}\theta_{-} & \frac{1}{2}(1+\xi_{-}\theta_{-}) & 0 \\
        -\Delta^*_{+}\theta_{+} & 0 & 0 & \frac{1}{2}(1+\xi_{+}\theta_{+}) \\
    \end{array} \right)
\end{equation}
and
\begin{equation}
 \underline{\xi}^b_\vk = \left( \begin{array}{cccc}
        \xi_{+} & 0 & 0 & \Delta_{+} \\
        0 & \xi_{-} & -\Delta_{-} & 0 \\
        0 & -\Delta^*_{-} & -\xi_{-} & 0 \\
        \Delta^*_{+} & 0 & 0 & -\xi_{+} \\
    \end{array} \right) \; .
\end{equation}
The deviations from equilibrium can be parameterized as follows:
\begin{eqnarray}
 \delta\underline{n}^b_\vk &=& \underline{U}^\dagger_{\vk+}\delta\underline{n}_\vk\underline{U}_{\vk-} = \left( \begin{array}{cccc}
        \delta n^b_{+} & 0 & 0 & \delta g^b_{+} \\
        0 & \delta n^b_{-} & -\delta g^b_{-} & 0 \\
        0 & -\delta g^{b*}_{-} & -\delta n^b_{-} & 0 \\
        \delta g^{b*}_{+} & 0 & 0 & -\delta n^b_{+}\\
    \end{array} \right) \label{eq:transp:def:nk} \\
 \delta\underline{\xi}^b_\vk &=& \underline{U}^\dagger_{\vk+}\delta\underline{\xi}_\vk\underline{U}_{\vk-} = \left( \begin{array}{cccc}
        \delta \xi^b_{+} & 0 & 0 & \delta\Delta^b_{+} \\
        0 & \delta \xi^b_{-} & -\delta\Delta^b_{-} & 0 \\
        0 & -\delta\Delta^{b*}_{-} & -\delta \xi^b_{-} & 0 \\
        \delta\Delta^{b*}_{+} & 0 & 0 & -\delta \xi^b_{+}\\
    \end{array} \right) \; . \label{eq:transp:def:xik}
\end{eqnarray}
Thus, we have now derived a set of equations in spin-- and band--basis [Eqs.~(\ref{eq:MKE:spin}) and (\ref{eq:MKE:band})] that allow us to determine the diagonal and off--diagonal non--equilibrium momentum distribution functions. In sections~\ref{subsec:Results:superfl_density}--\ref{subsec:Results:Raman} we will use these distribution functions to determine the normal and superfluid density, specific heat capacity, and the Raman response of NCS.
From now on, we will omit the index ``$b$'' indicating the band--basis, since all further considerations will be made in the band--picture.

\section{Solution by Bogoliubov transformation}
\label{sec:solution}
In what follows we will solve the kinetic equation~(\ref{eq:MKE:band}), derived in the previous section. For this purpose, we perform first a Bogoliubov transformation into quasiparticle space, where the kinetic equations are easily decoupled and then solved. For the subsequent inverse Bogoliubov transformation we will introduce parity projected quantities to obtain finally a relation between the diagonal and off--diagonal energy--shifts on the one side and the non--equilibrium distribution functions on the other.
As a fist step towards the solution of the kinetic equations, the momentum distribution matrix $\underline{n}_\vk$ and the energy matrix $\underline{\xi}_\vk$ (both in band--basis) are diagonalized through the following Bogoliubov transformation
\begin{eqnarray}
    \underline{\nu}_\vk &=& \underline{B}^\dagger_\vk\underline{n}_\vk\underline{B}_\vk = \left( \begin{array}{cccc}
        f(E_{+}) & 0 & 0 & 0 \\
        0 & f(E_{-}) & 0 & 0 \\
        0 & 0 & f(-E_{-}) & 0 \\
        0 & 0 & 0 & f(-E_{+}) \\
    \end{array} \right) \\
    \underline{E}_\vk &=& \underline{B}^\dagger_\vk\underline{\xi}_\vk\underline{B}_\vk = \left( \begin{array}{cccc}
        E_{+} & 0 & 0 & 0 \\
        0 & E_{-} & 0 & 0 \\
        0 & 0 & -E_{-} & 0 \\
        0 & 0 & 0 & -E_{+} \\
    \end{array} \right)
\end{eqnarray}
with the Fermi--Dirac distribution function $f(E_{\lambda})=[\exp(E_{\lambda}/\kB T)+1]^{-1}$.
The Bogoliubov matrix has been found to read in the band basis
\begin{equation}
    \underline{B}_\vk = \left( \begin{array}{cccc}
        u_{+} & 0 & 0 & v_{+} \\
        0 & u_{-} & -v_{-} & 0 \\
        0 & v^*_{-} & u_{-} & 0 \\
        -v^*_{+} & 0 & 0 & u_{+} \\
    \end{array} \right)
\end{equation}
with the coherence factors
\begin{eqnarray}
    u_{\lambda}(\vk) &=& \sqrt{\frac{1}{2}\left( 1+\frac{\xi_{\lambda}(\vk)}{E_{\lambda}(\vk)} \right)} \\
    v_{\lambda}(\vk) &=& -\sqrt{\frac{1}{2}\left( 1-\frac{\xi_{\lambda}(\vk)}{E_{\lambda}(\vk)} \right)}\frac{\Delta_{\lambda}(\vk)}{|\Delta_{\lambda}(\vk)|}
\end{eqnarray}
satisfying the condition $|u_{\lambda}|^2+|v_{\lambda}|^2=1$, by which the fermionic character of the Bogoliubov quasiparticles is established.
In order to solve the transport equation in the band basis (\ref{eq:MKE:band}), one may multiply from the left with the Bogoliubov matrix $\underline{B}^\dagger_{\vk+}$ and from the right with $\underline{B}_{\vk-}$. The result is
\begin{eqnarray}
 \hbar\omega\underline{B}^\dagger_{\vk+}\delta\underline{n}_\vk\underline{B}_{\vk-}
 &+& \underline{B}^\dagger_{\vk+}\delta\underline{n}_\vk \underline{B}_{\vk-}\underline{B}^\dagger_{\vk-} \underline{\xi}^0_{\vk-} \underline{B}_{\vk-}
   - \underline{B}^\dagger_{\vk+}\underline{\xi}^0_{\vk+} \underline{B}_{\vk+}\underline{B}^\dagger_{\vk+} \delta\underline{n}_\vk \underline{B}_{\vk-} \nn\\
 &=& \underline{B}^\dagger_{\vk+}\delta\underline{\xi}_\vk \underline{B}_{\vk-}\underline{B}^\dagger_{\vk-} \underline{n}^0_{\vk-} \underline{B}_{\vk-}
   - \underline{B}^\dagger_{\vk+}\underline{n}^0_{\vk+} \underline{B}_{\vk+}\underline{B}^\dagger_{\vk+} \delta\underline{\xi}_\vk \underline{B}_{\vk-}
\end{eqnarray}
or, more simply
\begin{equation}
 \hbar\omega\delta\underline{\nu}_\vk + \delta\underline{\nu}_\vk\underline{E}_{\vk-} - \underline{E}_{\vk+}\delta\underline{\nu}_\vk = \delta\underline{E}_\vk \underline{\nu}_{\vk-} - \underline{\nu}_{\vk+}\delta\underline{E}_\vk \; . \label{eq:MKE:QP}
\end{equation}
The new Bogoliubov--transformed quantities describing the deviation from equilibrium are identified from the preceding equations and labeled as follows:
\begin{eqnarray}
    \delta\underline{\nu}(\vk) &=& \underline{B}^\dagger_{\vk+}\delta\underline{n}_\vk\underline{B}_{\vk-}
    = \left( \begin{array}{cccc}
        \delta\nu_+(\vk) & 0 & 0 & \delta\gamma_+(\vk) \\
        0 & \delta\nu_-(\vk) & -\delta\gamma_-(\vk) & 0 \\
        0 & -\delta\gamma^*_-(\vk) & -\delta\nu_-(-\vk) & 0 \\
        \delta\gamma^*_+(\vk) & 0 & 0 & -\delta\nu_+(-\vk) \\
    \end{array} \right) \\
    \delta\underline{E}(\vk) &=& \underline{B}^\dagger_{\vk+} \delta\underline{\xi}_\vk \underline{B}_{\vk-}
    = \left( \begin{array}{cccc}
        \delta E_+(\vk) & 0 & 0 & \delta D_+(\vk) \\
        0 & \delta E_-(\vk) & -\delta D_-(\vk) & 0 \\
        0 & -\delta D^*_-(\vk) & -\delta E_-(-\vk) & 0 \\
        \delta D^*_+(\vk) & 0 & 0 & -\delta E_+(-\vk) \\
    \end{array} \right) \; .
\end{eqnarray}
The solution of Eq.~(\ref{eq:MKE:QP}) for the quasiparticle distribution functions is the set of the following eight equations ($\lambda=\pm$):
\begin{subequations} \label{eq:Solution:dnu}
\begin{eqnarray}
    \delta\nu_\lambda(\vk) &=& \frac{\eta^-_\lambda(\vk)}{\omega-\eta^-_\lambda(\vk)} \tilde y_\lambda(\vk) \delta E_\lambda(\vk) \; , \\
    \delta\nu_\lambda(-\vk) &=& -\frac{\eta^-_\lambda(\vk)}{\omega+\eta^-_\lambda(\vk)} \tilde y_\lambda(\vk) \delta E_\lambda(-\vk) \; , \\
    \delta\gamma_\lambda(\vk) &=& \frac{\eta^+_\lambda(\vk)}{\omega-\eta^+_\lambda(\vk)} \Theta_\lambda(\vk) \delta D_\lambda(\vk) \; , \\
    \delta\gamma^*_\lambda(\vk) &=& -\frac{\eta^+_\lambda(\vk)}{\omega+\eta^+_\lambda(\vk)} \Theta_\lambda(\vk) \delta D^*_\lambda(\vk) \; ,
\end{eqnarray}
\end{subequations}
where we have introduced the following abbreviations:
\begin{eqnarray}
    \eta^{\pm}_\lambda(\vk) &=& E_\lambda(\vk+)\pm E_\lambda(\vk-) \; , \\
    \tilde y_\lambda(\vk) &=& -\frac{f[E_\lambda(\vk+)]-f[E_\lambda(\vk-)]}{E_\lambda(\vk+)-E_\lambda(\vk-)} \; ,
\end{eqnarray}
and
\begin{equation}
    \Theta_\lambda(\vk) = \frac{1-f[E_\lambda(\vk+)]-f[E_\lambda(\vk-)]}{E_\lambda(\vk+)+E_\lambda(\vk-)} \; .
\end{equation}
The expressions for these quantities in the long--wavelength limit can be found in appendix~1.
In this limit, the difference quotient $\tilde y_\lambda(\vk)$ is equal to the Yosida kernel $y_\lambda(\vk)$ which is given by the derivative of the quasiparticle distribution function
\begin{equation}
 y_\lambda(\vk) = - \frac{\partial f[E_\lambda(\vk)]}{\partial E_\lambda(\vk)}= \frac{1}{4\kB T} \frac{1}{\cosh^2\left( \frac{E_\lambda(\vk)}{2\kB T} \right)}
\end{equation}
and is crucial for the temperature dependence of all response and transport functions.
Accordingly, $\Theta_\lambda(\vk)\stackrel{\vq\to0}{\rightarrow}\theta_\lambda(\vk)$ represents the kernel of the self--consistency equation~(\ref{Eq_gap_eq}).
It is instructive to note that the distribution functions $\delta\nu_\lambda(\vk)$ and $\delta\gamma_\lambda(\vk)$ have a clear physical meaning: The diagonal component $\delta\nu_\lambda(\vk)=\delta\langle \hat\alpha_\lambda^\dagger \hat\alpha_\lambda \rangle(\vk)$ describes the response of the Bogoliubov quasiparticles (with the quasiparticle creation and annihilation operators $\hat\alpha_\lambda^\dagger$, $\hat\alpha_\lambda$ in the band $\lambda$).
The off--diagonal component $\delta\gamma_\lambda(\vk)=\delta\langle \hat\alpha_\lambda \hat\alpha_\lambda \rangle(\vk)$ describes the pair--response.
Note that the abbreviations $\eta^\pm_\lambda(\vk)$ are of even ($+$) and odd ($-$) parity w.r.t. $\vk\rightarrow -\vk$ and become very simple expressions in the small wavelength limit (see appendix~1).

For the inverse Bogoliubov transformation it is convenient to introduce parity--projected quantities which are labeled by $s=\pm 1$:
\begin{eqnarray}
    \delta n^{(s)}(\vk) &=& \frac{1}{2}\left[ \delta n(\vk) + s\delta n(-\vk) \right] \; , \\
    \delta\xi^{(s)}(\vk) &=& \frac{1}{2}\left[ \delta\xi(\vk) + s\delta\xi(-\vk) \right] \; .
\end{eqnarray}
In almost the same manner also the off--diagonal components are decomposed by
\begin{equation}
    \delta g^{(s)}(\vk) = \frac{1}{2}\left[ \delta g(\vk)\frac{\Delta^*(\vk)}{|\Delta(\vk)|} + s\frac{\Delta(\vk)}{|\Delta(\vk)|}\delta g(-\vk) \right] \; ,
\end{equation}
and
\begin{equation}
    \delta\Delta^{(s)}(\vk) = \frac{1}{2}\left[ \delta\Delta(\vk)\frac{\Delta^*(\vk)}{|\Delta(\vk)|} + s\frac{\Delta(\vk)}{|\Delta(\vk)|}\delta\Delta(-\vk) \right] \; . \label{Eq:parity:delta}
\end{equation}
We use the same symmetry classification for the Bogoliubov transformed quantities.
The physical meaning of $\delta\Delta_\lambda(\vk,\vq,\omega)$ becomes clear after a decomposition into its real and imaginary part
\begin{eqnarray}
    \delta\Delta(\vk,\vq,\omega) &=& a(\vk,\vq,\omega)e^{i\varphi(\vq,\omega)} - \Delta(\vk) \\
    &=& \left[ \delta a(\vk,\vq,\omega) + i\delta\phi(\vq,\omega)|\Delta(\vk)| \right] \frac{\Delta(\vk)}{|\Delta(\vk)|} \; . \nn
\end{eqnarray}
With Eq.~(\ref{Eq:parity:delta}) we can identify $\delta\Delta^{(+)}(\vk,\vq,\omega)=\delta a(\vk,\vq,\omega)$ as the amplitude fluctuations and $\delta\Delta^{(-)}(\vk,\vq,\omega)/\Delta(\vk)=i\delta\varphi(\vq,\omega)$ as the phase fluctuations of the order parameter.

The off--diagonal energy shift $\delta\Delta^{(s)}_\lambda(\vk)$ can be determined from a straightforward variation of the self--consistency equation~(\ref{Eq_gap_eq}):
\begin{equation}
    \delta\Delta^{(s)}_\lambda(\vk) = \sum\limits_{\vk^\prime\mu} V^{\lambda\mu}_{\vk\vk^\prime} \delta\vg^{(s)}_\mu(\vk^\prime) \label{eq:solution:BT:sce}
\end{equation}
with $\delta g^{(s)}_\lambda(\vk)=-\theta_\lambda(\vk)\delta\Delta^{(s)}_\lambda(\vk)$. This off--diagonal self--consistency equation will play an important role for the gauge invariance of the theory, as will be discussed in section~\ref{subsec:Results:gaugemode}.

From the symmetry--classification we can assign to each transport and response function (see Table~\ref{tab:pert}) the corresponding momentum distribution function $\delta n^{(+)}_\lambda(\vk)$ or $\delta n^{(-)}_\lambda(\vk)$: The vertex function of the (charge) density-- and Raman--response is even in $\vk$, thus only the even distribution function $\delta n^{(+)}_\lambda(\vk)$ contributes to those response--functions. For the current--response (dynamic conductivity), the vertex--function ($a_\sigma(\vk)=e\vv_\vk$) is odd in momentum. Thus, only $\delta n^{(-)}_\lambda(\vk)$ contributes to the conductivity upon summation over $\vk$.
Furthermore, the Bogoliubov--transformation can now be written in this simple form
\begin{eqnarray}
    \left( \begin{array}{c} \delta\nu^{(s)}_\lambda(\vk) \\ \delta\gamma^{(s)}_\lambda(\vk) \end{array} \right) &=&
    \left( \begin{array}{cc}
            q^{(s)}_\lambda(\vk) & p^{(s)}_\lambda(\vk) \\
        -p^{(s)}_\lambda(\vk) & q^{(s)}_\lambda(\vk)
           \end{array} \right) \cdot
    \left( \begin{array}{c} \delta n^{(s)}_\lambda(\vk) \\ \delta g^{(s)}_\lambda(\vk) \end{array} \right)  \\
    \left( \begin{array}{c} \delta E^{(s)}_\lambda(\vk) \\ \delta D^{(s)}_\lambda(\vk) \end{array} \right) &=&
    \left( \begin{array}{cc}
            q^{(s)}_\lambda(\vk) & p^{(s)}_\lambda(\vk) \\
        -p^{(s)}_\lambda(\vk) & q^{(s)}_\lambda(\vk)
           \end{array} \right) \cdot
    \left( \begin{array}{c} \delta\xi^{(s)}_\lambda(\vk) \\ \delta\Delta^{(s)}_\lambda(\vk) \end{array} \right) \label{eq:BT:coherencefactors}
\end{eqnarray}
which might easily be inverted by using the sum rule
\begin{equation}
    \left[ q^{(s)}_\lambda(\vk) \right]^2 + \left[ p^{(s)}_\lambda(\vk) \right]^2 = 1 \;.
\end{equation}
Here, we have defined the real--valued coherence--factors
\begin{equation}
    q^{(s)}_\lambda(\vk) = |u_\lambda(\vk+)u_\lambda(\vk-)| - s|v_\lambda(\vk+)v_\lambda(\vk-)|
\end{equation}
and
\begin{equation}
    q^{(s)}_\lambda(\vk) = |u_\lambda(\vk+)v_\lambda(\vk-)| + s|u_\lambda(\vk-)v_\lambda(\vk+)|
\end{equation}
with the explicit form
\begin{equation}
    q^{(s)}_\lambda(\vk) = \sqrt{\frac{1}{2} + \frac{\xi_\lambda(\vk+)\xi_\lambda(\vk-) - s|\Delta_\lambda(\vk)|^2}{2E_\lambda(\vk+)E_\lambda(\vk-)}}
\end{equation}
and
\begin{equation}
    p^{(s)}_\lambda(\vk) = \sqrt{\frac{1}{2} - \frac{\xi_\lambda(\vk+)\xi_\lambda(\vk-) - s|\Delta_\lambda(\vk)|^2}{2E_\lambda(\vk+)E_\lambda(\vk-)}} \; .
\end{equation}
From Eqs.~(\ref{eq:BT:coherencefactors}) and (\ref{eq:MKE:QP}) we finally obtain the following solution of the matrix--kinetic equation
\begin{equation}
    \left( \begin{array}{c} \delta n^{+}_\lambda(\vk) \\ \delta n^{-}_\lambda(\vk) \\ \delta g^{+}_\lambda(\vk) \\ \delta g^{-}_\lambda(\vk) \end{array} \right) =
    \left( \begin{array}{cccc}
        N_{11} & N_{12} & N_{13} & N_{14} \\
        N_{21} & N_{22} & N_{23} & N_{24} \\
        N_{31} & N_{32} & N_{33} & N_{34} \\
        N_{41} & N_{42} & N_{43} & N_{44}
    \end{array} \right) \cdot
    \left( \begin{array}{c} \delta \xi^{+}_\lambda(\vk) \\ \delta \xi^{-}_\lambda(\vk) \\ \delta \Delta^{+}_\lambda(\vk) \\ \delta \Delta^{-}_\lambda(\vk) \end{array} \right) \label{eq:BT:solution}
\end{equation}
The vector on the left hand side contains the non--equilibrium momentum distribution functions [defined in Eq.~(\ref{eq:transp:def:nk})] which can be expressed in terms of the diagonal and off--diagonal energy--shifts [defined in Eq.~(\ref{eq:transp:def:xik}) and obtained from Table~\ref{tab:pert} and Eq.~(\ref{eq:solution:BT:sce})].
The matrix--elements $N_{ij}$ read in detail:
\begin{subequations} \label{eq:BT:solution:BT}
\begin{align}
    N_{11} &= q^{(+)2}_\lambda(\vk) \tilde y^{(+)}_\lambda(\vk) + p^{(+)2}_\lambda(\vk) \Theta^{(+)}_\lambda(\vk) \\
    N_{12} &= q^{(+)}_\lambda(\vk) q^{(-)}_\lambda(\vk) \tilde y^{(-)}_\lambda(\vk) + p^{(+)}_\lambda(\vk) p^{(-)}_\lambda(\vk) \Theta^{(-)}_\lambda(\vk)  \\
    N_{13} &= q^{(+)}_\lambda(\vk) p^{(+)}_\lambda(\vk) \left[ \tilde y^{(+)}_\lambda(\vk) - \Theta^{(+)}_\lambda(\vk) \right]  \\
    N_{14} &= q^{(+)}_\lambda(\vk) p^{(-)}_\lambda(\vk) \tilde y^{(-)}_\lambda(\vk) - q^{(-)}_\lambda(\vk) p^{(+)}_\lambda(\vk) \Theta^{(-)}_\lambda(\vk)  \\
\nn \\
    N_{22} &= q^{(-)2}_\lambda(\vk) \tilde y^{(+)}_\lambda(\vk) + p^{(-)2}_\lambda(\vk) \Theta^{(+)}_\lambda(\vk)  \\
    N_{23} &= q^{(-)}_\lambda(\vk) p^{(+)}_\lambda(\vk) \tilde y^{(-)}_\lambda(\vk) - q^{(+)}_\lambda(\vk) p^{(-)}_\lambda(\vk) \Theta^{(-)}_\lambda(\vk)  \\
    N_{24} &= q^{(-)}_\lambda(\vk) p^{(-)}_\lambda(\vk) \left[ \tilde y^{(+)}_\lambda(\vk) - \Theta^{(+)}_\lambda(\vk) \right]  \\
\nn \\
    N_{33} &= p^{(+)2}_\lambda(\vk) \tilde y^{(+)}_\lambda(\vk) + q^{(+)2}_\lambda(\vk) \Theta^{(+)}_\lambda(\vk)  \\
    N_{34} &= p^{(+)}_\lambda(\vk) p^{(-)}_\lambda(\vk) \tilde y^{(-)}_\lambda(\vk) + q^{(+)}_\lambda(\vk) q^{(-)}_\lambda(\vk) \Theta^{(-)}_\lambda(\vk)  \\
\nn \\
    N_{44} &= p^{(-)2}_\lambda(\vk) \tilde y^{(+)}_\lambda(\vk) + q^{(-)2}_\lambda(\vk) \Theta^{(+)}_\lambda(\vk) \; .
\end{align}
\end{subequations}
The matrix--elements $N_{ij}$ are symmetric, i.e. $N_{ij}=N_{ji}$ and the occurring products of coherence--factors can be found in the appendix~1. Above, we have defined the following abbreviations:
\begin{eqnarray}
    \tilde y^{(s)}_\lambda(\vk) &=& \frac{\eta^{(s)2}_\lambda(\vk)}{\omega^2-\eta^{(s)2}_\lambda(\vk)} \tilde y_\lambda(\vk) \\
    \Theta^{(s)}_\lambda(\vk) &=& \frac{\eta^{(s)2}_\lambda(\vk)}{\omega^2-\eta^{(s)2}_\lambda(\vk)} \Theta_\lambda(\vk) \; . \nn
\end{eqnarray}
The matrix--elements $N_{13}$, $N_{23}$ and $N_{34}$ are shown to be odd w.r.t. $\xi_\lambda(\vk)\rightarrow-\xi_\lambda(\vk)$. Thus in a particle--hole symmetric theory, these terms will vanish upon integration over $\xi_\lambda(\vk)$ and are labeled $O(\mbox{pha})$ which stands for ``particle--hole asymmetric''.
It is convenient to rewrite these matrix elements in terms of the functions
\begin{eqnarray}
    \lambda_\lambda(\vk) &=& \left[ p^{(+)2}_\lambda(\vk) - q^{(-)2}_\lambda(\vk) \right] \left[ \tilde y^{(+)}_\lambda(\vk) - \Theta^{(+)}_\lambda(\vk) \right] \\
    \Phi_\lambda(\vk) &=& q^{(+)2}_\lambda \tilde y_\lambda(\vk) + p^{(+)2}_\lambda \tilde y_\lambda(\vk)\Theta_\lambda(\vk)  \\
    \frac{\Theta^{(+)}_\lambda(\vk)}{2} &=& \frac{\eta^{(+)2}_\lambda(\vk) \Theta_\lambda(\vk) - \eta^{(-)2}_\lambda(\vk) \tilde y_\lambda(\vk)}{\eta^{(+)2}_\lambda(\vk) - \eta^{(-)2}_\lambda(\vk)}
\end{eqnarray}
where the first one, $\lambda_\lambda(\vk)$ is referred to as the Tsuneto--function \cite{Tsuneto:1960:01}.
A straightforward but lengthy calculation yields
\begin{subequations} \label{eq:BT:solution:ME}
\begin{align}
    N_{11} &= \frac{\eta^2 \Phi_\lambda(\vk) - \omega^2\lambda_\lambda(\vk)}{\omega^2-\eta^2} \\
    N_{12} &= \frac{\omega \eta [\Phi_\lambda(\vk) - \lambda_\lambda(\vk)]}{\omega^2-\eta^2}  \\
    N_{13} &= O(\mbox{\rm pha})  \\
    N_{14} &= \frac{\omega}{2\Delta_\lambda(\vk)}\lambda_\lambda(\vk)  \\
\nn \\
    N_{22} &= \frac{\eta^2 [\Phi_\lambda(\vk) - \lambda_\lambda(\vk)]}{\omega^2-\eta^2}  \\
    N_{23} &= O(\mbox{\rm pha})  \\
    N_{24} &= \frac{\eta}{2\Delta_\lambda(\vk)}\lambda_\lambda(\vk)  \\
\nn \\
    N_{33} &= -\frac{\theta^{(+)}_\lambda(\vk)}{2} - \frac{\omega^2-\eta^2 - 4\Delta^2_\lambda(\vk)}{4\Delta^2_\lambda(\vk)}\lambda_\lambda(\vk)  \\
    N_{34} &= O(\mbox{\rm pha})  \\
\nn \\
    N_{44} &= -\frac{\theta^{(+)}_\lambda(\vk)}{2} - \frac{\omega^2-\eta^2}{4\Delta^2_\lambda(\vk)}\lambda_\lambda(\vk) \; ,
\end{align}
\end{subequations}
where $\eta=\vv_\vk\cdot\vq$.
Note, that all expressions are valid in the whole quasiclassical limit, i.e. for $\vq\ll\kF$ and $\hbar\omega\ll\EF$.
For small wave numbers, as required e.g. in the Raman case, the Tsuneto and related functions $\lambda_\lambda(\vk)$, $\Phi_\lambda(\vk)$ and $\theta_\lambda^{(+)}$ simplify considerably. The results for such a small--$\vq$ expansion can be found in appendix~1.
Our further considerations for response and transport properties require both main results of this section:
The solution of the transport equation in quasiparticle space, given by Eq.~(\ref{eq:Solution:dnu}), will be used directly in section~\ref{subsec:Results:heatcapacity} to derive the specific heat capacity in NCS (see Table~\ref{tab:pert}).
While for the discussion of the gauge mode (section~\ref{subsec:Results:gaugemode}), the normal and superfluid density (section~\ref{subsec:Results:superfl_density}) and the Raman response (section~\ref{subsec:Results:Raman}) the non--equilibrium distribution functions after an inverse Bogoliubov transformation, given in Eq.~(\ref{eq:BT:solution}) and Eq.~(\ref{eq:BT:solution:ME}), are necessary.

\section{Gauge invariance}
\label{subsec:Results:gaugemode}
The gauge invariance of our theory is an important issue, which will be discussed in the following section. Therefore, we determine the gauge modes and insert them into the transport equations. An integration of these transport equations yields a continuity equation which demonstrates the gauge invariance of our theory for $\hbar\omega\ll\EF$ and $\vq\ll\kF$.
For this purpose, it is very instructive to rebuild the original distribution function by combining $\delta n^+_\lambda$ and $\delta n^-_\lambda$ from Eq.~(\ref{eq:BT:solution}) and Eq.~(\ref{eq:BT:solution:ME}):
\begin{equation}
\omega\delta n_\lambda-\eta \left[ \delta n_\lambda +\Phi_\lambda\delta\xi_\lambda \right] = -\lambda_\lambda \left[ \omega\delta\xi^+_\lambda + \eta\delta\xi^-_\lambda \right] + \lambda_\lambda\left(\omega^2-\eta^2\right)\frac{\delta\Delta^-_\lambda}{2\Delta_\lambda} \; . \label{eq:gm:BE}
\end{equation}
The left hand side of this equation is of the same structure as the linearized Landau--Boltzmann equation of the normal state. In what follows, we want to
discuss the right hand side of the above equation. Note that all terms coupling to $\delta\Delta^+_\lambda$ have vanished because of particle--hole symmetry. This means that the amplitude fluctuations of the order parameter do not contribute to the response in a particle--hole symmetric theory. The phase fluctuations are also given by Eq.~(\ref{eq:BT:solution}):
\begin{equation}
 \delta g^-_\lambda + \left[ \frac{\theta^+_\lambda}{2} + \frac{\omega^2-\eta^2}{4\Delta^2_\lambda}\lambda_\lambda \right]\delta\Delta^-_\lambda = \frac{\omega\delta\xi^+_\lambda+\eta\delta\xi^-_\lambda}{2\Delta_\lambda}\lambda_\lambda \; .
\end{equation}
Multiplication with the pairing--interaction $V^{\lambda\mu}_{\vk\vk^\prime}$ and summation over $\vk^\prime$ and the band--index $\mu$ yields
\begin{eqnarray}
 \delta\Delta^-_\lambda(\vk) &+& \sum\limits_{\vk^\prime\mu} V^{\lambda\mu}_{\vk\vk^\prime} \left[ \theta_\mu +\delta\theta_\mu + \frac{\omega^2-\eta^2}{4\Delta^2_\mu}\lambda_\mu \right]\delta\Delta^-_\mu(\vk^\prime) \\
 &=& \sum\limits_{\vk^\prime\mu} V^{\lambda\mu}_{\vk\vk^\prime} \frac{\omega\delta\xi^+_\mu(\vk^\prime)+\eta\delta\xi^-_\mu(\vk^\prime)}{2\Delta_\mu(\vk^\prime)}\lambda_\mu(\vk^\prime) \; , \nn
\end{eqnarray}
where we have introduced $\delta\theta_\lambda=\theta^+_\lambda/2-\theta_\lambda$. It can be shown, that the $\xi_\mu(\vk)$--integral over $\delta\theta_\mu$ vanishes identically for all $\vq$. Using the equilibrium gap--equation [Eq.~(\ref{Eq_gap_eq})] we arrive at
\begin{equation}
 \sum\limits_\mu \frac{\delta\Delta^-_\mu}{|\Delta_\mu|} \sum\limits_{\vk^\prime} V^{\lambda\mu}_{\vk\vk^\prime} \frac{\omega^2-\eta^2}{4|\Delta_\mu(\vk^\prime)|}\lambda_\mu(\vk^\prime) = \sum\limits_{\vk^\prime\mu} V^{\lambda\mu}_{\vk\vk^\prime} \frac{\omega\delta\xi^+_\mu(\vk^\prime)+\eta\delta\xi^-_\mu(\vk^\prime)}{2\Delta_\mu(\vk^\prime)}\lambda_\mu(\vk^\prime) \; .
\end{equation}
These are two coupled equations (for $\lambda=\pm$) which determine the phase fluctuations of the order parameter (gauge mode).
Note that in the weak coupling BCS theory, there are only two collective excitations possible: 
the Anderson--Bogoliubov and 2$\Delta$ mode. In NCS, there exist two gauge modes due to the band 
splitting, which can be connected  with the particle number conservation law. 
In addition, due to existence of a triplet fraction, there could be further collective excitation analogous to 
Leggett's SBSOS~\cite{Leggett:1975:01} modes predicted for the superfluid phases of $^3$He. The latter should be connectable with the spin conservation law in NCS. Finally, massive collective modes with frequencies below 2$\Delta/\hbar$ may exist in NCS.
It can be shown, that the right hand side of Eq.~(\ref{eq:gm:BE}) vanishes upon $\vk$ and $\lambda$ (band) summation when inserting the above expressions for the gauge mode. This leads us to the following continuity equation for the electron density:
\begin{equation}
 \omega\sum\limits_{\vk,\lambda} \delta n_\lambda(\vk) - \vq\cdot\sum\limits_{\vk,\lambda} \vv_\vk \left[ \delta n_\lambda(\vk) + \Phi_\lambda(\vk)\delta\xi_\lambda(\vk) \right] = 0 \; .
\end{equation}
For a conserved quantity such as the particle or charge density $a_\vk=1,\; e$ we can identify the corresponding generalized density and current density
\begin{eqnarray}
 \delta n_a &=& \sum\limits_{\vk,\lambda} a_\vk \delta n_\lambda(\vk) \label{eq:Solution:density}\\
 \vj_a      &=& \sum\limits_{\vk,\lambda} a_\vk \vv_\vk \left[ \delta n_\lambda(\vk) + \Phi_\lambda(\vk)\delta\xi_\lambda(\vk) \right] \; , \label{eq:Solution:current}
\end{eqnarray}
obeying the continuity equation
\begin{equation}
 \omega\delta n_a - \vq\cdot\vj_a=0 \; .
\end{equation}
Therefore, we have demonstrated charge conservation and gauge invariance of the theory for $\hbar\omega\ll\EF$ and $\vq\ll\kF$.

\section{Normal and superfluid density}
\label{subsec:Results:superfl_density}
The normal and superfluid density are derived in the static and long--wavelength limit ($\omega\rightarrow 0$ and $\vq\rightarrow 0$). In order to preserve gauge invariance, gradient terms of the order $O(\vq)$ are still taken into account. The parity--projected distribution functions are obtained from Eq.~(\ref{eq:BT:solution}) and from Eq.~(\ref{eq:BT:solution:ME}):
\begin{eqnarray}
 \delta n^+_\lambda(\vk) &=& -\phi_\lambda(\vk)\delta\xi^+_\lambda(\vk) \\
 \delta n^-_\lambda(\vk) &=& -\left[\phi_\lambda(\vk)-\lambda_\lambda(\vk)\right]\delta\xi^-_\lambda(\vk) + \eta\lambda_\lambda(\vk)\frac{\delta\Delta^-_\lambda(\vk)}{2\Delta_\lambda(\vk)} \; ,
\end{eqnarray}
where we made use of the $\vq\rightarrow 0$ limit with the coherence--factors $q^-_\lambda(\vk)\rightarrow 1$, $p^-_\lambda(\vk)\rightarrow 0$, and $\Phi_\lambda(\vk)\rightarrow \phi_\lambda(\vk)$, $\tilde y_\lambda(\vk)\rightarrow y_\lambda(\vk)$, as well as the Tsuneto--function $\lambda_\lambda(\vk)\rightarrow \phi_\lambda(\vk) - y_\lambda(\vk)$ (see appendix~1). The combined expression for $\delta n^+_\lambda(\vk)$ and $\delta n^-_\lambda(\vk)$ are now inserted in Eq.~(\ref{eq:Solution:current}) to derive the supercurrent density (vertex--function $a_\vk=e$):
\begin{eqnarray}
    \vj^s_i &=& \sum\limits_{\vp \lambda} e\, \vv_{\vp i} \left[ \delta n_\lambda(\vp) + \phi_\lambda(\vp)\delta\xi^-_\lambda(\vp) \right] \\
    &=& e\sum\limits_{\vp \lambda} \vv_{\vp i} \vv_{\vp j} \lambda_\lambda(\vp) \left( -\frac{e}{c}\vA + \frac{\hbar}{2}\nabla\delta\varphi_\lambda \right) \; . \nn
\end{eqnarray}
Here we used the result from section~\ref{sec:solution} that $\delta\Delta^-_\lambda(\vk)/\Delta_\lambda(\vk)=i\delta\varphi_\lambda$ represents the phase fluctuations of the order parameter. These phase fluctuations ensure gauge invariance in the above expression for the supercurrent. By rewriting the supercurrent as product of the superfluid density and the corresponding velocity $\vv^s$, we can easily identify
\begin{eqnarray}
    \vj^s &=& e\; \vn^s\cdot\vv^s \\
    \vv^s &=& \frac{e}{m}\left( -\frac{e}{c}\vA + \frac{\hbar}{2}\nabla\delta\varphi_\lambda \right) \; .
\end{eqnarray}
Therefore, the superfluid and normal fluid density tensor read
\begin{eqnarray}
    n^s_{ij} &=& \sum\limits_{\vp \lambda} \vp_i \vv_{\vp j} \lambda_\lambda(\vp) \\
    n^n_{ij} &=& n\delta_{ij} - \vn^s_{ij} = \sum\limits_{\vp \lambda} \vp_i \vv_{\vp j} y_\lambda(\vp) \; .
\end{eqnarray}
Thus, in this static and small--$\vq$--limit we obtain a very clear picture: The Yosida--kernel $y_\lambda(\vk) = - \partial f[E_\lambda(\vk)]/\partial E_\lambda(\vk)$ generates the normal fluid density and the Tsuneto--function $\lambda_\lambda(\vk)$ gives rise to the superfluid density.

It is important to realize, that this result can be derived in the following alternative simple way from local equilibrium considerations.
In terms of the Fermi--Dirac distribution function on both bands $f[E_\lambda(\vp)]$ for the Bogoliubov quasiparticles, the supercurrent can be written in the standard quantum--mechanical form:
\begin{eqnarray}
j^{\rm s}_i&=&nv^{\rm s}_i+\frac{1}{V}\sum_{\vp \lambda}
v_{\vp i}(\vp) f(E_{\lambda}(\vp)+\vp\cdot\vv^{\rm s})\\
&=&nv^{\rm s}_i+\frac{1}{V}\sum_{\vp \lambda}v_{\vp i}
\left\{f(E_\lambda(\vp))+\frac{\partial f(E_\lambda(\vp))}{\partial E_\lambda(\vp)}p_jv^{\rm s}_j\right\} \nn \\
&=&\left\{n\delta_{ij}-\frac{1}{V}\sum_{\vp \lambda}\frac{p_i}{m}
\left(-\frac{\partial f(E_\lambda(\vp))}{\partial E_\lambda(\vp)}\right)p_j\right\}v^{\rm s}_j \; . \nn
\end{eqnarray}
This immediately implies the definition of the normal fluid density in the form
\begin{eqnarray}
n^{\rm n}_{ij}&=&\frac{1}{V}\sum_{\vp \lambda} p_iv_j y_\lambda(\vp) \; . 
\end{eqnarray}
Thus, the results obtained with our simple local equilibrium picture
are in agreement with the results in Ref.~\cite{Einzel:2003:01}.

\section{The specific heat capacity}
\label{subsec:Results:heatcapacity}
In order to derive the specific heat capacity, we start from an expression for the entropy of a NCS, which has to be written in the general form
\begin{eqnarray}
T\sigma(T)&=&-\frac{\kB}{V}\sum_{\vp\lambda}
f[E_\lambda(\vp)]\ln f[E_\lambda(\vp)]+\{1-f[E_\lambda(\vp)]\}\ln \{1-f[E_\lambda(\vp)]\} \nn \\
&=&\frac{1}{V}\sum_{\vp \lambda}\xi^2_\lambda(\vp) y_{\vp}^{(\lambda)} \; .
\end{eqnarray}
The change of the entropy as a consequence of a temperature change $\delta T$ can then
be written in the form \cite{Einzel:2003:01}
\begin{eqnarray}
T\delta\sigma(T)&=&\frac{1}{V}\sum_{\vp \lambda}E_\lambda(\vp)\delta \nu_\lambda(\vp) \; ,
\end{eqnarray}
where the quasiparticle distribution function is given by Eq.~(\ref{eq:Solution:dnu}). In the static and homogenous limit, i.e. $\omega\rightarrow 0$ and $\vq\rightarrow 0$, this expression simplifies considerably to $\delta\nu_\lambda(\vk)=y_\lambda(\vk)\delta E_\lambda(\vk)$. The quasiparticle energy shift for a temperature change is $\delta E_\lambda(\vk)=(E_\lambda(\vk)/T - \partial E_\lambda(\vk)/\partial T)\delta T$ for each band \cite{Einzel:2003:01}. Therefore, our result for the entropy change reads
\begin{eqnarray}
T\delta\sigma(T)&=&\frac{1}{V}\sum_{\vp \lambda}y_\lambda(\vp) E_\lambda(\vp)\left[E_\lambda(\vp)-T\frac{\partial E_\lambda(\vp)}{\partial T}\right]\delta T \nn \\
&=&C_V(T)\delta T
\end{eqnarray}
and one may easily identify the specific heat capacity as
\begin{eqnarray}
C_V(T) &=& \frac{1}{V}\sum_{\vp \lambda}y_\lambda(\vp) \left[E^2_\lambda(\vp)-\frac{T}{2}\frac{\partial \Delta^2_\lambda(\vp)}{\partial T}\right] \; .
\end{eqnarray}
%
%
An alternative way to derive the specific heat capacity employs again the concept of local equilibrium:
\begin{eqnarray}
T\delta\sigma(T)&=&\frac{1}{V}\sum_{\vp \lambda}E_\lambda(\vp)\delta f(E_\lambda(\vp)) \; .
\end{eqnarray}
The change of the BQP (Fermi--Dirac) distribution function with temperature has two causes:
first the direct change $T\to T+\delta T$ and second the change of the BQP
energy with temperature through the $T$--dependence of the energy gap:
\begin{eqnarray}
\delta f(E_\lambda(\vp))&=&f\left(\frac{E_\lambda(\vp)+\frac{\partial E_\lambda(\vp)}{\partial T}\delta T}{\kB[T+\delta T]}\right)
-f\left(\frac{E_\lambda(\vp)}{\kB T}\right) \\
&=&\underbrace{\left(-\frac{\partial f(E_\lambda(\vp))}{\partial E_\lambda(\vp)}\right)}_{y_\lambda(\vp)}\left(\frac{E^{0}_{\vp \lambda}}{T}-\frac{\partial E_\lambda(\vp)}{\partial T}\right)\delta T \; . \nn
\end{eqnarray}
Hence we arrive to the same result for the entropy change
\begin{eqnarray}
T\delta\sigma(T)&=&\frac{1}{V}\sum_{\vp \lambda}y_\lambda(\vp) E_\lambda(\vp)\left[E_\lambda(\vp)-T\frac{\partial E_\lambda(\vp)}{\partial T}\right]\delta T\\
&=&C_V(T)\delta T \nn
\end{eqnarray}
and the result for the specific heat capacity is confirmed.
Again, like in the case of the normal and superfluid density, the result for the specific heat capacity
can be viewed to consist of contributions from the two bands, in the sense that the sum over the
spin projections $\sigma=\pm 1$ is replaced by a sum over the pseudospin variable $\lambda=\pm$.

\section{A case study: Raman response}
\label{subsec:Results:Raman}
In the following section we will discuss in detail the electronic Raman response for $T=0$ in NCS \cite{Klam:2008:02}. An extensive description of the electronic Raman effect in unconventional superconductors can be found in Ref.~\cite{Devereaux:1995:01}.
A Raman experiment detects the intensity of the scattered light with frequency--shift $\omega=\omega_I-\omega_S$, where the incoming photon of frequency $\omega_I$ is scattered on an elementary excitation and gives rise to a scattered photon with frequency $\omega_S$ and a momentum transfer $\vq$. The differential photon scattering cross section of this process is given by Ref.~\cite{Klein:1984:01}
\begin{equation}
    \frac{\partial^2\sigma}{\partial\omega\partial\Omega} = \frac{\omega_S}{\omega_I}r^2_0 S_{\gamma\gamma} (\vq,\omega)
\end{equation}
with the solid angle $\Omega$ and the Thompson radius $r_0=e^2/mc^2$. The generalized structure function $S_{\gamma\gamma}(\vq,\omega)$ is connected through the fluctuation--dissipation theorem to the imaginary part of the Raman response function $\chi_{\gamma\gamma}(\vq,\omega)$:
\begin{equation}
    S_{\gamma\gamma}(\vq,\omega) = - \frac{\hbar}{\pi} \left[ 1+ n(\omega) \right] \chi^{\prime\prime}_{\gamma\gamma}(\vq,\omega) \; .
\end{equation}
Here, $n(\omega)=\left[ \exp (\hbar\omega/\kB T) -1 \right]$ denotes the Bose distribution.
After Coulomb renormalisation and in the long--wavelength limit ($\vq=0$), the Raman response function is given by the imaginary part of (see also Ref.~\cite{Monien:1990:01})
\begin{eqnarray}
 \chi_{\gamma\gamma}(\omega) &=& \chi_{\gamma\gamma}^{(0)}(\omega) - \frac{\left[\chi^{(0)}_{\gamma 1}(\omega)\right]^2}{\chi^{(0)}_{11}(\omega)} \; . \label{Eq_Raman_01}
\end{eqnarray}
Within our notation, the unscreened Raman response is given by
\begin{equation}
    \chi^{(0)}_{ab}(\omega) = \frac{1}{V}\sum\limits_{\vp,\sigma} a_\vp b_\vp \lambda_\vp(\omega) \; ,
\end{equation}
where the vertex--functions $a_\vp$, $b_\vp$ are either $1$ or the corresponding momentum--dependent Raman vertex $\gamma\equiv\gamma^{(R)}_\vk$ that describes the coupling of polarized light to the sample.
The long--wavelength limit of the Tsuneto--function $\lambda_\vp(\vq=0)= 4\Delta^2_\vp \theta_\vp/$  $(4E^2_\vp-\omega^2)$ is given in appendix~1 and since we are interested in the $T=0$ Raman response it is possible to perform the integration on the energy variable $\xi_\vk$ (see e.g. \cite{Devereaux:1995:01}).
Note that the second term in Eq.~(\ref{Eq_Raman_01}) is often referred to as the screening contribution that originates from gauge invariance. Since the ASOC leads to a splitting of the Fermi surface, the total Raman response is given by $\chi_{\gamma\gamma}^{\rm total} = \sum_{\lambda=\pm} \chi_{\gamma\gamma}^\lambda$ with $\chi^\pm_{\gamma\gamma}=\chi_{\gamma\gamma}(\Delta_\pm)$, in which the usual summation over the spin variable $\sigma$ is replaced by a summation over the pseudo--spin (band) index $\lambda$. With Eq.~(\ref{Eq_gap}) the unscreened Raman response for both bands in the clean limit [$l\gg\xi(0)$ with the mean free path $l$ and the coherence length $\xi(T=0)$] can be analytically expressed as
\begin{eqnarray}
  \Im\chi_{\gamma\gamma}^{(0)\pm} = \frac{\pi\NF^\pm\psi}{\omega} \Re \left\langle \gamma^{(R)\,2}_\vk \frac{\left| 1\pm p|\vgamma_\vk| \right|^2}{\sqrt{(\frac{\omega}{2\psi})^2 - \left| 1\pm p|\vgamma_\vk| \right|^2}} \right\rangle_{\rm FS} \; .
\end{eqnarray}
Here, $\NF^\pm$ reflect the different densities of states on both bands and $\langle\ldots\rangle_{\rm FS}$ denotes an average over the Fermi surface.
We consider small momentum transfers ($\vq\rightarrow 0$) and neglect interband scattering processes, assuming non--resonant scattering. Then, the Raman tensor is approximately given by
\begin{equation}
 \vgamma^{(R)}_\vk = m \sum\limits_{i,j} \hat\ve^S_i \frac{\partial^2 \epsilon(\vk)}{\hbar^2\partial k_i \partial k_j} \hat\ve^I_j \; ,
\end{equation}
where $\hat\ve^{S,I}$ denote the unit vectors of scattered and incident polarization light, respectively.
The light polarization selects elements of this Raman tensor, where $\vgamma^{(R)}_\vk$ can be decomposed into its symmetry components and, after a straight forward calculation (see appendix~2), expanded into a set of basis functions on a spherical Fermi surface.
Our results for the tetragonal group C$_{4v}$ are
\begin{subequations} \label{Eq_vertex_C4v}
\begin{align}
    \gamma^{(R)}_{A_1} &= \sum\limits_{k=0}^\infty \sum\limits_{l=0}^{l\leq k/2} \gamma^{(R)}_{k,l}\cos4l\phi\, \sin^{2k}\theta , \\
    \gamma^{(R)}_{B_1} &= \sum\limits_{k=1}^\infty \sum\limits_{l=1}^{l\leq (k+1)/2} \gamma^{(R)}_{k,l}\cos(4l-2)\phi\, \sin^{2k}\theta ,  \\
    \gamma^{(R)}_{B_2} &= \sum\limits_{k=1}^\infty \sum\limits_{l=1}^{l\leq (k+1)/2} \gamma^{(R)}_{k,l}\sin(4l-2)\phi\, \sin^{2k}\theta ,
\end{align}
\end{subequations}
and for the cubic group $O$ we obtain
\begin{subequations} \label{Eq_vertex_O432}
\begin{align}
    \gamma^{(R)}_{A_1} &= \sum\limits_{k=0}^\infty \sum\limits_{l=0}^{l\leq k/2} \gamma^{(R)}_{k,l}\cos4l\phi\, \sin^{2k}\theta , \\
    \gamma^{(R)}_{E^{(1)}} &= \gamma^{(R)}_0 (2-3\sin^2\theta) + \ldots ,  \\
    \gamma^{(R)}_{E^{(2)}} &= \sum\limits_{k=1}^\infty \sum\limits_{l=1}^{l\leq (k+1)/2} \gamma^{(R)}_{k,l}\cos(4l-2)\phi\, \sin^{2k}\theta ,  \\
    \gamma^{(R)}_{T_2} &= \sum\limits_{k=1}^\infty \sum\limits_{l=1}^{l\leq (k+1)/2} \gamma^{(R)}_{k,l}\sin(4l-2)\phi\, \sin^{2k}\theta
\end{align}
\end{subequations}
in a backscattering--geometry experiment ($z\overline{z}$) \footnote{The vertices E$^{(1)}$ and E$^{(2)}$ seem to be quite different, but it turns out that the Raman response is exactly the same because E$^{(1)}$ and E$^{(2)}$ are both elements of the same symmetry class.}. In what follows, we neglect higher harmonics and thus use only the leading term in the expansions of $\vgamma^{(R)}_\vk$ \footnote{Due to screening, the constant term (${k=0, l=0}$) in the A$_1$ vertex generates no Raman response, thus we used (${k=1, l=0}$). For all the other vertices the leading term is given by (${k=1, l=1}$).}.

\begin{figure}[t] 
\begin{minipage}[b]{0.05\linewidth}
  \rotatebox{90}{\fontfamily{phv}\selectfont \hspace{1cm} \large{Raman intensity (arbitrary units)}}
\end{minipage}
\begin{minipage}[b]{0.4\linewidth}
 \includegraphics[angle=0, width=1.0\linewidth]{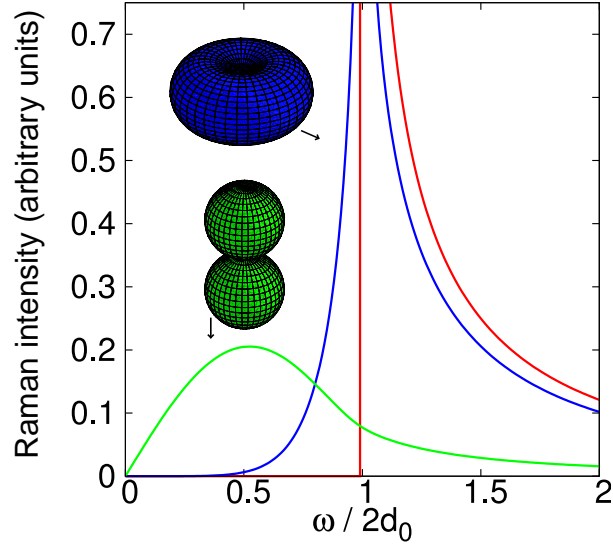}
\end{minipage}
\caption{Calculated Raman spectra for a pure triplet order parameter (i.e. $\psi=0$) for B$_{1,2}$ polarization of the point group C$_{4v}$ in backscattering geometry ($z\overline{z}$). The ABM (axial) state with $|\vd_\vk|=d_0 \sin \theta$ is displayed in blue and the polar state with $|\vd_\vk|=d_0 |\cos \theta|$ in green. For a comparison, also the threshold--behavior of the Raman response for the BW state (red) with $|\vd_\vk|=d_0$ is shown.}
\label{fig:triplet}
\end{figure}

In general, due to the mixing of a singlet and a triplet component to the superconducting condensate, one expects a two--peak structure in parity--violated NCS, reflecting both pair--breaking peaks for the linear combination [see Eq.~(\ref{Eq_gap})] of the singlet order parameter $\psi_\vk$ (extensively discussed in Ref.~\cite{Devereaux:1995:01}) and the triplet order parameter $\vd_\vk$ (shown in Fig.~\ref{fig:triplet}), respectively. The ratio $p=d/\psi$, however, is unknown for both types of ASOCs.

%
%

How does the Raman spectra look like for a pure triplet p--wave state?
Some representative examples, see Fig.~\ref{fig:triplet}, are the Balian--Werthamer (BW) state, the Anderson--Brinkman--Morel (ABM or axial) state, and the polar state. The simple pseudoisotropic BW state with $\vd_\vk= d_0 \hat\vk$ [equivalent to Eq.~(\ref{Eq_gk_cubic}) for $\mathrm{g}_3=0$], as well as previous work on triplet superconductors, restricted on a (cylindrical) 2D Fermi surface, generates the same Raman response as an $s$--wave superconductor \cite{Kee:2003:01}. However, in three dimensions we obtain more interesting results for the axial state with $\vd_\vk=d_0(\hat k_y\hat\ve_x - \hat k_x\hat\ve_y)$ [equivalent to Eq.~(\ref{Eq_gk_tetragonal}) for $\mathrm{g}_\parallel=0$]. The Raman response for this axial state in B$_1$ and B$_2$ polarizations for $\mathcal{G}=C_{4v}$ is given by
\begin{eqnarray}
 \chi^{\prime\prime}_{B_{1,2}} (x) &=& \frac{\pi \NF \gamma^{(R)\,2}_0}{128} \label{Eq_triplet_ABM} \\
 &\times& \left( -10 -\frac{28}{3}x^2-10x^4 + \frac{5+3x^2+3x^4+5x^6}{x} \ln \left| \frac{x+1}{x-1} \right| \right) \nn
\end{eqnarray}
with the dimensionless frequency $x=\omega/2d_0$. An expansion for low frequencies reveals a characteristic exponent [$\chi^{\prime\prime}_{B_{1,2}} \propto \left(\omega/2d_0\right)^6$], which is due to the overlap between the gap and the vertex function. 
Moreover, we calculate the Raman response for the polar state with $\vd_\vk=d_0\hat k_z \hat\ve_x$; in this case one equatorial line node crosses the Fermi surface and we obtain:
\begin{align}
 \chi^{\prime\prime}_{B_{1,2}} (x) = \frac{\pi \NF \gamma^{(R)\,2}_0}{8x}
 \begin{cases}
    \frac{\pi}{2}x^2-\frac{3\pi}{4}x^4+\frac{5\pi}{16}x^6 & x\leq 1 \\
    \left( x^2-\frac{3}{2}x^4+\frac{5}{8}x^6 \right) \arcsin \frac{1}{x} & x > 1\\
      - \left( \frac{1}{3}-\frac{13}{12}x^2+\frac{5}{8}x^4 \right) \sqrt{x^2-1}
 \end{cases}
\end{align}
with the trivial low frequency expansion $\chi^{\prime\prime}_{B_{1,2}} \propto \omega/2d_0$.
While the pair--breaking peaks for the BW and ABM state were both located at $\omega=2d_0$ (similar to the B$_{1g}$ polarization in the singlet $d$--wave case, which is peaked at $2\Delta_0$), for the polar state this peak is significantly shifted to lower frequencies ($\omega=1.38 d_0$).

\begin{figure}[t] 
\begin{minipage}[c]{0.05\linewidth}
  \rotatebox{90}{\fontfamily{phv}\selectfont \large{Raman intensity (arbitrary units)}}
\end{minipage}
\begin{minipage}[c]{0.9\linewidth}
  \includegraphics[angle=0, width=1.0\linewidth]{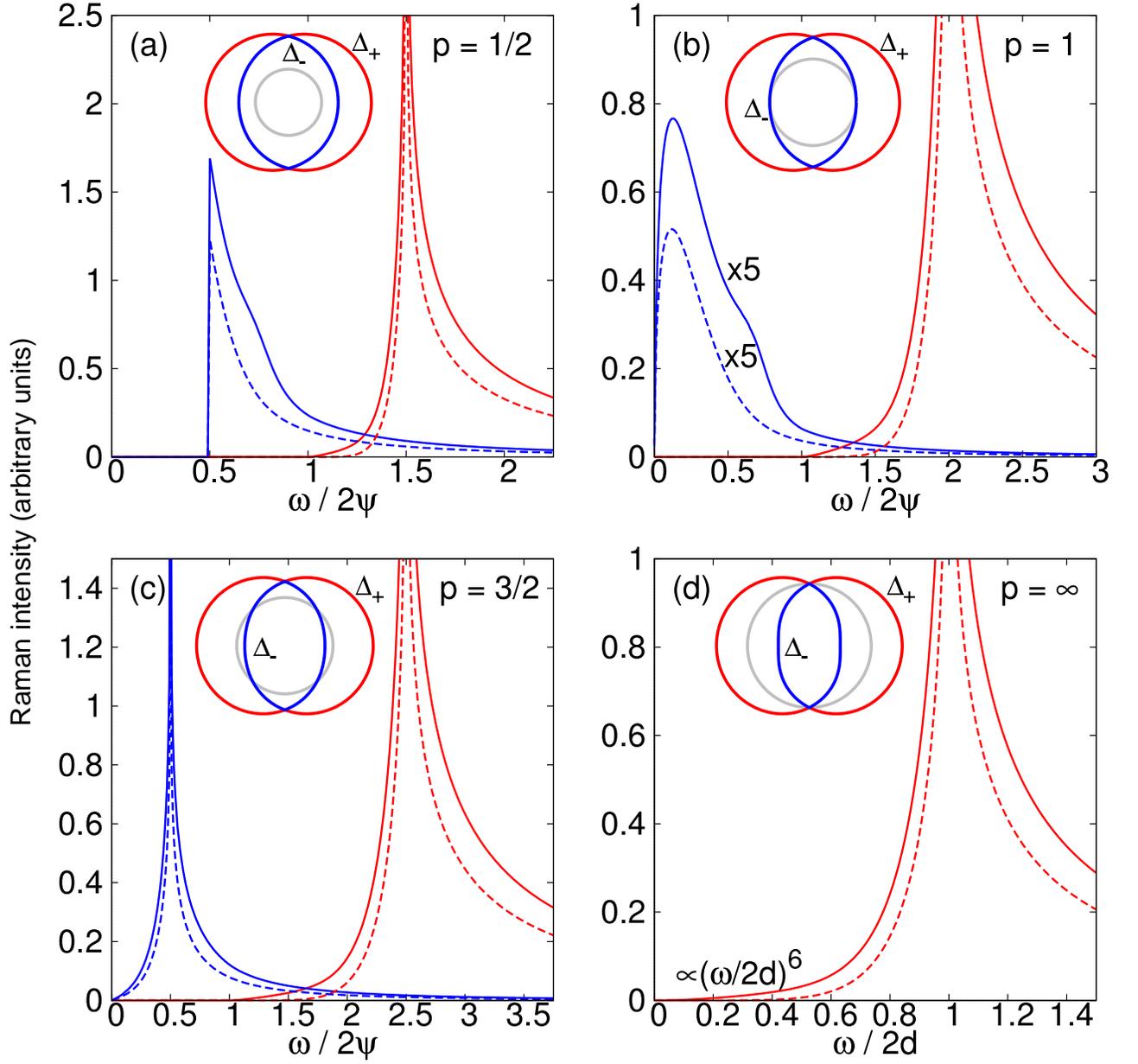}
\end{minipage}
\caption{Calculated Raman spectra $\chi^{\prime\prime}_{\gamma\gamma}(\Delta_-)$ [blue] and $\chi^{\prime\prime}_{\gamma\gamma}(\Delta_+)$ [red] for A$_1$ (solid lines) and for B$_{1,2}$ (dashed lines) polarizations for the point group C$_{4v}$. We obtain the same spectra for the B$_1$ and B$_2$ symmetry. The polar diagrams in the insets demonstrate the four qualitative different cases for the unknown ratio $p=d/\psi$.}
\label{fig:C4v}
\end{figure}

Let's turn to the predicted Raman spectra for the tetragonal point group $\mathcal{G}=C_{4v}$.
In Fig.~\ref{fig:C4v} we show the calculated Raman response using Eq.~(\ref{Eq_gk_tetragonal}) with $\mathrm{g}_\parallel=0$. This Rashba--type of ASOC splits the Fermi surface into two bands; while on the one band the gap function is $\Delta_{\vk} = \psi \left( 1 + p|\vgamma_\vk| \right)\equiv\Delta_+$, it is $\Delta_- \equiv \psi \left( 1 - p|\vgamma_\vk| \right)$ on the other band.
Thus, depending on the ratio $p=d/\psi$, four different cases (see polar diagrams in the insets) have to be considered: (a) no nodes; (b) one (equatorial) line node ($\Delta_-$ band); (c) two line nodes ($\Delta_-$ band); and (d) two point nodes on both bands.
Since the Raman intensity in NCS is proportional to the imaginary part of
\begin{equation}
\chi_{\gamma\gamma}^{\rm total}=\chi_{\gamma\gamma}(\Delta_-)+\chi_{\gamma\gamma}(\Delta_+) \; ,
\end{equation}
it is interesting to display both contributions separately (blue and red lines, respectively).
Although (except for $\psi=0$) we always find two pair--breaking peaks at
\begin{equation}
\frac{\omega}{2\psi}=|1\pm p|
\end{equation}
 we stress that our results for NCS are not just a superposition of a singlet and a triplet spectra. This is clearly demonstrated in Fig.~\ref{fig:C4v}(a), for example, in which we show the results for a small triplet contribution ($p=1/2$). For $\chi_{\gamma\gamma}^{\prime\prime}(\Delta_-)$ we find a threshold behavior with an adjacent maximum value of $\chi^{\prime\prime}_{B_{1,2}}(\Delta_-)=\NF^-\gamma^{(R)\,2}_0\,\pi^2/8\sqrt{p^{-1}-1}$. In contrast for $\chi_{\gamma\gamma}^{\prime\prime}(\Delta_+)$ a zero Raman signal to twice the singlet contribution followed by a smooth increase and a singularity is obtained \footnote{Note that even though the gap function does not depend on $\phi$ (see Fig.~\ref{fig:gk}), we obtain a small polarization--dependence. This unusual behavior only in A$_1$ symmetry is due to screening and leads to a small shoulder for $p\leq 1$.}.
In the special case, in which the singlet contribution equals the triplet one ($p=1$), the gap function $\Delta_-$ displays an equatorial line node without sign change. This is displayed in Fig.~\ref{fig:C4v}b. Because of this nodal structure and strong weight from the vertex function ($\propto\sin^2\theta$), many low energy quasiparticles can be excited, which leads to this square--root--like increase in the Raman intensity. In this special case the pair--breaking peak is located very close to elastic scattering ($\omega=0.24\psi$).
In Fig.~\ref{fig:C4v}(c) the gap function $\Delta_-$ displays two circular line nodes. The corresponding Raman response for $p>1$ shows two singularities with different low frequency power laws [$\chi^{\prime\prime}_{B_{1,2}}(\Delta_-)\propto\omega/2\psi$ and $\chi^{\prime\prime}_{B_{1,2}}(\Delta_+)\propto(\omega/2\psi-1)^{11/2}$].
Finally, for $p\gg1$ one recovers the pure triplet cases (d) which is given analytically by Eq.~(\ref{Eq_triplet_ABM}).

\begin{figure}[t] 
\begin{minipage}[c]{0.05\linewidth}
  \rotatebox{90}{\fontfamily{phv}\selectfont \large{Raman intensity (arbitrary units)}}
\end{minipage}
\begin{minipage}[c]{0.9\linewidth}
     \includegraphics[angle=0, width=1.0\linewidth]{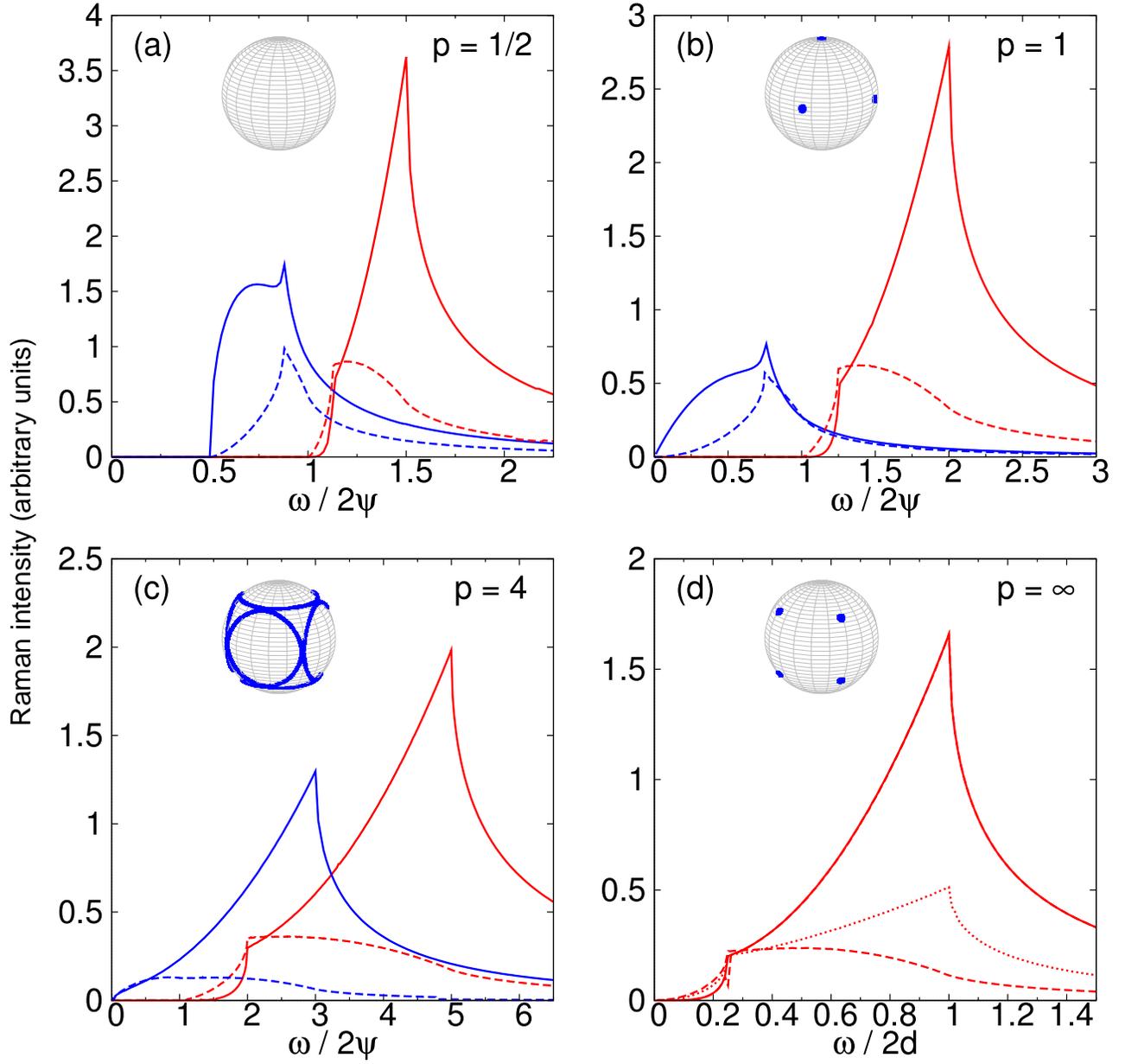}
\end{minipage}
\caption{Calculated Raman spectra $\chi_{\gamma\gamma}(\Delta_-)$ [blue] and $\chi_{\gamma\gamma}(\Delta_+)$ [red] for E (solid lines), T$_2$ (dashed lines) and A$_1$ [dotted line, only in (d)] polarizations for the point group $O$. The insets display the point and line nodes of the gap function $\Delta_-$.}
\label{fig:O432}
\end{figure}

The Raman response for the point group $O$, using Eq.~(\ref{Eq_gk_cubic}), is shown in Fig.~\ref{fig:O432}.
We again consider four different cases: (a) no nodes; (b) six point nodes ($\Delta_-$ band); (c) six connected line nodes ($\Delta_-$ band); and (d) 8 point nodes (both bands) as illustrated in the insets.
Obviously, the pronounced angular dependence of $|\vgamma_\vk|$ leads to a strong polarization--dependence. Thus we get different peak positions for the E and T$_2$ polarizations in $\chi_{\gamma\gamma}^{\prime\prime}(\Delta_+)$.
As a further consequence, the Raman spectra reveals up to two kinks on each band ($+$,$-$) at
\begin{equation}
\frac{\omega}{2\psi}=|1\pm p/4|
\end{equation}
and
\begin{equation}
\frac{\omega}{2\psi}=|1\pm p| \; .
\end{equation}
Interestingly, the T$_2$ symmetry displays only a change in slope at $\omega/2\psi=|1+p|$ instead of a kink.
Furthermore, no singularities are present.
Nevertheless, the main feature, namely the two--peak structure, is still present and one can directly deduce the value of $p$ from the peak and kink positions.
Finally, for $p\gg1$ one recovers the pure triplet case (d), in which the unscreened Raman response is given by
\begin{eqnarray}
  \chi^{\prime\prime}_{\gamma\gamma}(\omega) \propto \frac{2d}{\omega} \Re \left\langle \gamma^{(R)\,2}_\vk \frac{|\vgamma_\vk|^2}{\sqrt{(\omega/2d+|\vgamma_\vk|)(\omega/2d-|\vgamma_\vk|)}} \right\rangle_{\rm FS}\;.
\end{eqnarray}
Clearly, only the area on the Fermi surface with ${\omega/2d>|\vgamma_\vk|}$ contributes to the Raman intensity. Since $|\vgamma_\vk|\in [0,1]$ has a saddle point at $|\vgamma_\vk|=1/4$, we find kinks at characteristic frequencies $\omega/2d=1/4$ and $\omega/2d=1$.
In contrast to the Rashba--type ASOC, we find a characteristic low energy expansion $\propto(\omega/2d)^2$ for both the A$_1$ and E symmetry, while $\propto(\omega/2d)^4$ for the T$_2$ symmetry.
Assuming weak coupling BCS theory, we expect the pair--breaking peaks (as shown in Fig.~\ref{fig:O432}) for Li$_2$Pd$_x$Pt$_{3-x}$B roughly in the range $4\,$cm$^{-1}$ to $30\,$cm$^{-1}$.

\section{Conclusion}
\label{sec:Conclusion}
In this chapter, we derived response and transport functions for noncentrosymmetric superconductors from a kinetic  theory with particular emphasis on the Raman response.
We started from the generalized von Neumann equation which describes the evolution of the momentum distribution function in time and space and derived a linearized matrix--kinetic (Boltzmann) equation in $\omega$--$\vq$--space. This kinetic equation is a $4\times 4$ matrix equation in both particle--hole (Nambu) and spin space. We explored the Nambu--structure and solved the kinetic equation quite generally by first performing an SU(2) rotation into the band--basis and second applying a Bogoliubov--transformation into quasiparticle space. Our theory is particle--hole symmetric, applies to any kind of antisymmetric spin--orbit coupling, and holds for arbitrary quasiclassical frequency and momentum with $\hbar\omega\ll\EF$ and $|\vq|\ll\kF$.
Furthermore, assuming a separable ansatz in the pairing interaction, we demonstrated gauge invariance and charge conservation for our theory.
Within this framework, we derived expressions for the normal and superfluid density and compared the results in the static and long--wavelength limit with those from a local equilibrium analysis. The same investigations were done for the specific heat capacity. In both cases we recover the same results, which validates our theory.

Finally, we presented analytic and numeric results for the electronic (pair--breaking) Raman response in noncentrosymmetric superconductors for zero temperature. Therefore we analyzed the two most interesting classes of tetragonal and cubic symmetry, applying for example to CePt$_3$Si ($\mathcal{G}=C_{4v}$) and Li$_2$Pd$_x$Pt$_{3-x}$B ($\mathcal{G}=O$). Accounting for the antisymmetric spin--orbit coupling, we provide various analytic results such as the Raman vertices for both point groups, the Raman response for several pure triplet states, and power laws and kink positions for mixed--parity states. Our numeric results cover all relevant cases from weak to strong triplet--singlet ratio and demonstrate a characteristic two--peak structure for Raman spectra of non--centrosymmetric superconductors. Our theoretical predictions can be used to analyze the underlying condensate in parity--violated noncentrosymmetric superconductors and allow the determination of the unknown triplet--singlet ratio.

\begin{acknowledgements}
We thank M. Sigrist for helpful discussions.
\end{acknowledgements}

\section*{Appendix 1: Small $\vq$--expansion}
\addcontentsline{toc}{section}{Appendix 2}
For small wave numbers, i.e. $\vq\rightarrow 0$, the Tsuneto and related functions, which play an important role in the matrix--elements $N_{ij}$ [see Eq.~(\ref{eq:BT:solution:ME})], will simplify considerably. Taking into account terms to the order $O(\eta_\vk^2)$ with $\eta_\vk=\vv_\vk\cdot\vq$, we obtain the well--known expression for the Tsuneto--function \cite{Hirschfeld:1989:01}
\begin{equation}
 \lim\limits_{\vq\rightarrow 0}\lambda_\lambda(\vk) = - 4\Delta^2_\lambda(\vk) \frac{(\omega^2-\eta_\vk^2)\theta_\lambda(\vk)+\eta_\vk^2\phi_\lambda(\vk)}{\omega^2[\omega^2-4E^2_\lambda(\vk)]-\eta_\vk^2[\omega^2-4\xi^2_\lambda(\vk)]}
\end{equation}
where
\begin{equation}
 \phi_\lambda(\vk)=-\frac{\partial n_\lambda(\vk)}{\partial\xi_\lambda(\vk)}=\frac{\xi^2_\lambda(\vk)}{E^2_\lambda(\vk)}y_\lambda(\vk)+\frac{\Delta^2_\lambda(\vk)}{E^2_\lambda(\vk)}\theta_\lambda(\vk)
\end{equation}
is the derivative of the electron distribution function in the band $\lambda$ and
\begin{equation}
 y_\lambda(\vk) = - \frac{\partial f[E_\lambda(\vk)]}{\partial E_\lambda(\vk)}= \frac{1}{4\kB T} \frac{1}{\cosh^2\left( \frac{E_\lambda(\vk)}{2\kB T} \right)}
\end{equation}
is the derivative of the quasiparticle distribution function.

The following limits are also of interest: the homogenous limit ($\vq=0$), e.g. for the Raman response and the static limit ($\omega=0$), used in local equilibrium situation
\begin{eqnarray}
 \lambda_\lambda(\vk,\vq=0) &=& \frac{4\Delta^2_\lambda(\vk)\theta_\lambda(\vk)}{4E^2_\lambda(\vk)-\omega^2} \\
 \lim\limits_{\omega\rightarrow 0}\lim\limits_{\vq\rightarrow 0} \lambda_\lambda(\vk) &=& \phi_\lambda(\vk)-y_\lambda(\vk) \; .
\end{eqnarray}
For the following small $\vq$--expansion we omitted the band--label $\lambda$ for better readability:
\begin{subequations}
\begin{align}
\lim_{\vq\to 0}\theta_\vk^+&= 2\theta_\vk + \frac{\eta_\vk^2}{4E_\vk^2}\left[ \frac{\Delta_\vk^2-2\xi_\vk^2}{E_\vk^2}(y_\vk-\theta_\vk)
-\frac{\xi_\vk^2}{E_\vk}f_\vk^{\prime\prime}\right] \\
\lim_{\vq\to 0}\theta_\vk^-&= \frac{\eta_\vk\xi_\vk}{E_\vk^2}(y_\vk-\theta_\vk) \\
\lim_{\vq\to 0}\Phi_\vk&= \phi_\vk +\frac{\eta_\vk^2}{4E_\vk^2}
\frac{\Delta_\vk^2(\Delta_\vk^2-4\xi_\vk^2)}{2E_\vk^4}(y_\vk-\theta_\vk)-\frac{\eta_\vk^2\xi_\vk^2}{4E_\vk^2}\left[
\frac{\Delta_\vk^2}{E_\vk}f_\vk^{\prime\prime}+\frac{\xi_\vk^2}{6E_\vk^3}f_\vk^{\prime\prime\prime}\right] \\
\delta\theta_\vk&= \frac{\theta_\vk^+}{2}-\theta_\vk \nn \\
&= \frac{\eta_\vk^2}{8E_\vk^2}\left[\frac{\Delta_\vk^2-2\xi_\vk^2}{E_\vk^2}(y_\vk-\theta_\vk)
-\frac{\xi_\vk^2}{E_\vk}f_\vk^{\prime\prime}\right]  \\
\delta\phi_\vk&= \Phi_\vk-\phi_\vk \nn \\
&= \frac{\eta_\vk^2}{4E_\vk^2}\frac{\Delta_\vk^2(\Delta_\vk^2-4\xi_\vk^2)}{2E_\vk^4}(y_\vk-\theta_\vk) - \frac{\eta_\vk^2\xi_\vk^2}{4E_\vk^2}\left[
\frac{\Delta_\vk^2}{E_\vk}f_\vk^{\prime\prime}+\frac{\xi_\vk^2}{6E_\vk^3}f_\vk^{\prime\prime\prime}\right] \; ,
\end{align}
\end{subequations}
where $f_\vk^{(n)}$ denotes the nth derivative of $f(E_\vk)$ with respect to $E_\vk$.
Furthermore we find the following expansions:
\begin{subequations}
\begin{align}
\lim_{\vq\to 0}\eta^+_\vk&= 2E_\vk\left(1+\frac{\eta_\vk^2\Delta_\vk^2}{8E_\vk^4}\right) \\
\lim_{\vq\to 0}\eta^-_\vk&= \frac{\xi_\vk}{E_\vk}\eta_\vk\left(1-\frac{\eta_\vk^2\Delta_\vk^2}{8E_\vk^4}\right)  \\
\lim_{\vq\to 0}\tilde y_\vk&= y_\vk -\frac{\eta_\vk^2}{8E_\vk^2}\left(\frac{\Delta_\vk^2}{E_\vk}\nu_\vk^{\prime\prime}
+\frac{\xi_\vk^2}{3}\nu_\vk^{\prime\prime\prime}\right)  \\
\lim_{\vq\to 0}\Theta_\vk&= \theta_\vk +\frac{\eta_\vk^2}{8E_\vk^2}\left[ \frac{\Delta_\vk^2}{E_\vk^2}(y_\vk-\theta_\vk)
-\frac{\xi_\vk^2}{E_\vk}\nu_\vk^{\prime\prime}\right] \; .
\end{align}
\end{subequations}
The ten products of coherence--factors in Eq.~(\ref{eq:BT:solution:BT}) have the following explicit form:
\begin{subequations}
\begin{align}
    \left[ q^{(s)}_\vk \right]^2 &= \frac{1}{2} \frac{E_{\vk+}E_{\vk-} + \xi_{\vk+}\xi_{\vk-} - s\Delta^2_{\vk}}{E_{\vk+}E_{\vk-}} \\
    \left[ p^{(s)}_{\vk} \right]^2 &= \frac{1}{2} \frac{E_{\vk+}E_{\vk-} - \xi_{\vk+}\xi_{\vk-} + s\Delta^2_{\vk}}{E_{\vk+}E_{\vk-}}  \\
    q^{(+)}_{\vk} q^{(-)}_{\vk}  &= \frac{1}{2} \frac{E_{\vk-}\xi_{\vk+} + E_{\vk+}\xi_{\vk-}}{E_{\vk+}E_{\vk-}}  \\
    p^{(+)}_{\vk} p^{(-)}_{\vk}  &= \frac{1}{2} \frac{E_{\vk-}\xi_{\vk+} - E_{\vk+}\xi_{\vk-}}{E_{\vk+}E_{\vk-}}  \\
    q^{(s)}_{\vk} p^{(s)}_{\vk}   &= \frac{\Delta_{\vk}}{2} \frac{\xi_{\vk+} + s\xi_{\vk-}}{E_{\vk+}E_{\vk-}}  \\
    q^{(-)}_{\vk} p^{(+)}_{\vk}   &= \frac{\Delta_{\vk}}{2} \frac{E_{\vk+} + E_{\vk-}}{E_{\vk+}E_{\vk-}}
\end{align}
\end{subequations}
and the small $\vq$--limit of each coherence--factors reads:
\begin{subequations}
\begin{align}
    \lim_{\vq\rightarrow 0} q^{(+)}_\vk &= \frac{\xi_\vk}{E_\vk} \left( 1- \frac{\eta^2_\vk\Delta^2_\vk}{4E^4_\vk} \right) \\
    \lim_{\vq\rightarrow 0} q^{(-)}_\vk &= 1 - \frac{\eta^2_\vk\Delta^2_\vk}{8E^4_\vk}  \\
    \lim_{\vq\rightarrow 0} p^{(+)}_\vk &= \frac{\Delta_\vk}{E_\vk} \left( 1 + \frac{\eta^2_\vk\xi^2_\vk}{4E^4_\vk} \right)  \\
    \lim_{\vq\rightarrow 0} p^{(-)}_\vk &= \frac{\eta_\vk\Delta_\vk}{2E^2_\vk} \; .
\end{align}
\end{subequations}

\section*{Appendix 2: Derivation of the Raman vertices}
\addcontentsline{toc}{section}{Appendix 3}

In order to derive the relevant expressions for the polarization--dependent Raman vertices, we start from a general dispersion relation for tetragonal symmetry (C$_{4v}$)
\begin{align}
 \epsilon_\vk &= \sum_{n=1}^{\infty} \sum_{r=0}^{\infty} a_{n,r}^{C_{4v}} \left[ \cos(nk_xa)+\cos(nk_ya) \right]\cos(rk_zc) \\
 &+ \sum_{n=0}^{\infty} \sum_{r=0}^{\infty} b_{n,r}^{C_{4v}} \cos(nk_xa) \cos(nk_ya) \cos(rk_zc) \nn \\
 &+ \sum_{n=1}^{\infty} \sum_{m=1}^{\infty} \sum_{r=0}^{\infty} c_{n,m,r}^{C_{4v}} \left[ \cos(nk_xa) \cos(mk_ya) + \cos(mk_xa) \cos(nk_ya) \right]\cos(rk_zc) \nn
\end{align}
and for the cubic symmetry ($O$)
\begin{align}
 \epsilon_\vk &= \sum_{n=1}^{\infty} a_{n}^{O} \left[ \cos(nk_xa)+\cos(nk_ya) + \cos(nk_zc) \right] \\
 &+ \sum_{n=0}^{\infty} b_{n}^{O} \cos(nk_xa) \cos(nk_ya) \cos(rk_za) \nn \\
 &+ \sum_{n=1}^{\infty} \sum_{m=1}^{n-1} c_{n,m}^{O} \left[ \cos(mk_xa)\cos(mk_ya)\cos(nk_za) \right. \nn \\
 & + \left.\cos(mk_xa)\cos(nk_ya)\cos(mk_za) + \cos(nk_xa)\cos(mk_ya)\cos(mk_za) \right] \nn \\
 &+ \sum_{n=2}^{\infty} \sum_{m=1}^{n-1} \sum_{r=0}^{m-1} d_{n,m,r}^{O} \left[ \cos(nk_xa)\cos(mk_ya)\cos(rk_za) \right. \nn \\
 & + \cos(nk_xa)\cos(rk_ya)\cos(mk_za) + \cos(mk_xa)\cos(nk_ya)\cos(rk_za) \nn \\
 & + \cos(rk_xa)\cos(nk_ya)\cos(mk_za)  + \cos(mk_xa)\cos(rk_ya)\cos(nk_za) \nn \\
 & + \left.\cos(rk_xa)\cos(mk_ya)\cos(nk_za) \right] \; .\nn
\end{align}
Time reversal symmetry allows only for even functions of momentum $\vk$ in the energy dispersion. Furthermore the dispersion must be invariant under all symmetry elements of the point group $\mathcal{G}$ of the crystal.
For small momentum transfers and nonresonant scattering, the Raman tensor is given by the effective--mass approximation
\begin{equation}
  \vgamma (\vk) = m \sum\limits_{i,j} \hat\ve^S_i \frac{\partial^2 \epsilon(\vk)}{\hbar^2\partial k_i \partial k_j} \hat\ve^I_j \; .
\end{equation}
where $\hat\ve^{S,I}$ denote the unit vectors of the scattered and incident polarization light, respectively.\\
The light polarization vectors select elements of the Raman tensor according to
\begin{equation}
 \gamma^{IS}_\vk = \ve^I\cdot\vgamma^{(R)}_\vk\cdot\ve^S \; ,
\end{equation}
where the Raman tensor $\vgamma_\vk$ can be decomposed into its symmetry components and later expanded into Fermi surface harmonics:
\begin{align}
\vgamma_\vk^{C_{4v}} &= \left( \begin{array}{ccc}
                \gamma_{A_1^{(1)}} + \gamma_{B_1} & \gamma_{B_2} & \gamma_{E^{(1)}} \\
        \gamma_{B_2} & \gamma_{A_1^{(1)}} - \gamma_{B_1} & \gamma_{E^{(2)}} \\
        \gamma_{E^{(1)}} & \gamma_{E^{(2)}} & \gamma_{A_1^{(2)}} \\
              \end{array} \right) \\
 \vgamma_\vk^{O} &= \left( \begin{array}{ccc}
                \gamma_{A_1} + \gamma_{E^{(1)}} -\sqrt{3}\gamma_{E^{(2)}} & \gamma_{T_2^{(1)}} & \gamma_{T_2^{(2)}} \\
        \gamma_{T_2^{(1)}} & \gamma_{A_1} + \gamma_{E^{(1)}} +\sqrt{3}\gamma_{E^{(2)}} & \gamma_{T_2^{(3)}} \\
        \gamma_{T_2^{(2)}} & \gamma_{T_2^{(3)}} & \gamma_{A_1} - 2\gamma_{E^{(1)}} \\
              \end{array} \right) \; .
\end{align}
Here we have omitted all non--Raman active symmetries such as A$_{2g}$.
The vertices A$_1^{(1)}$ and A$_1^{(2)}$ are equal up to some constants determined by the band structure, and the vertices for $E^{(1)}$ and $E^{(2)}$ in C$_{4v}$ differ only by a rotation of the azimuthal angle $\phi$ by $\pi/2$. Since this rotation is an element of the corresponding point groups, these vertices are identical, too. The same holds for $T_2^{(1)}$, $T_2^{(2)}$ and $T_2^{(3)}$. Therefore the upper indices will be omitted in the following (whenever possible).
For the tetragonal group C$_{4v}$ the A$_1$, B$_1$, B$_2$ and E symmetries are Raman active in backscattering geometry.
Relevant polarizations for this group are:
\begin{align}
 \gamma_\vk^{xx} &= \gamma_\vk^{A_1} + \gamma_\vk^{B_1} & \gamma_\vk^{x^\prime x^\prime} &= \gamma_\vk^{A_1} + \gamma_\vk^{B_2} \nn\\
 \gamma_\vk^{yy} &= \gamma_\vk^{A_1} - \gamma_\vk^{B_1} & \gamma_\vk^{y^\prime y^\prime} &= \gamma_\vk^{A_1} - \gamma_\vk^{B_2} \nn\\
 \gamma_\vk^{xy} &= \gamma_\vk^{B_2} & \gamma_\vk^{x^\prime y^\prime} &= \gamma_\vk^{B_1} \\
 \gamma_\vk^{xz} &= \gamma_\vk^{E} & \gamma_\vk^{RR} &= \gamma_\vk^{A_1} \nn\\
 \gamma_\vk^{yz} &= \gamma_\vk^{E} & \gamma_\vk^{LL} &= \gamma_\vk^{A_1} \nn\\
 \gamma_\vk^{zz} &= \gamma_\vk^{A_1} & \gamma_\vk^{RL} &= \gamma_\vk^{B_1} - i\gamma_\vk^{B_2} \; . \nn
\end{align}
The cubic group $O$ reveals three Raman active symmetries, namely A$_1$, (E$^{(1)}$, E$^{(2)}$), and T$_2$ (still assuming backscattering geometry).
The relevant polarizations are:
\begin{align}
 \gamma_\vk^{xx} &= \gamma_\vk^{A_1} + \gamma_\vk^{E^{(1)}} - \sqrt{3}\gamma_\vk^{E^{(2)}} & \gamma_\vk^{x^\prime x^\prime} &= \gamma_\vk^{A_1} + \gamma_\vk^{E^{(1)}} + \gamma_\vk^{T_2} \nn\\
 \gamma_\vk^{yy} &= \gamma_\vk^{A_1} + \gamma_\vk^{E^{(1)}} + \sqrt{3}\gamma_\vk^{E^{(2)}} & \gamma_\vk^{y^\prime y^\prime} &= \gamma_\vk^{A_1} + \gamma_\vk^{E^{(1)}} - \gamma_\vk^{T_2} \nn\\
 \gamma_\vk^{xy} &= \gamma_\vk^{T_2} & \gamma_\vk^{x^\prime y^\prime} &= -\sqrt{3}\gamma_\vk^{E^{(2)}} \\
 \gamma_\vk^{xz} &= \gamma_\vk^{T_2} & \gamma_\vk^{RR} &= \gamma_\vk^{A_1}+ \gamma_\vk^{E^{(1)}} \nn\\
 \gamma_\vk^{yz} &= \gamma_\vk^{T_2} & \gamma_\vk^{LL} &= \gamma_\vk^{A_1}+ \gamma_\vk^{E^{(1)}} \nn\\
 \gamma_\vk^{zz} &= \gamma_\vk^{A_1} - 2\gamma_\vk^{E^{(1)}} & \gamma_\vk^{RL} &= -\sqrt{3}\gamma_\vk^{E^{(2)}} - i\gamma_\vk^{T_2} \; . \nn
\end{align}
Here, we have defined the unit polarization vectors $\hat \vx^\prime = (\hat\vx + \hat\vy)/\sqrt{2}$ and $\hat \vy^\prime = (\hat\vx - \hat\vy)/\sqrt{2}$. L and R denote left and right circularly polarized light with positive and negative helicity, respectively ($\ve^L = (\hat\vx+i\hat\vy)/\sqrt{2}$, $\ve^R = (\hat\vx-i\hat\vy)/\sqrt{2}$). Note that in a backscattering configuration the polarization vectors $\ve^{I,S}$ are pinned to the coordinate system of the crystal axes. Therefore some caution is advised when choosing the proper helicity for the scattered polarization vector $\ve^S$.
Although the Raman vertices E$^{(1)}$ and E$^{(2)}$ seem to look completely different, the Raman response turns out to be exactly the same. From a tight--binding analysis we obtain the same (band--structure) prefactors for both vertices, thus $\gamma_\vk^{E^{(1)}}$ and $\sqrt{3}\gamma_\vk^{E^{(2)}}$ generate both the same Raman response.
Note that it is not possible to measure A$_1$ and E$^{(1)}$ independently in backscattering geometry with the crystal c--axis aligned parallel to the laser beam.

The Raman vertices are extracted from the band structure by comparing the symmetry components of the Raman tensor with the second derivative of the energy dispersion. This can be done by solving a set of 6 coupled linear equations -- the 6 equations correspond exactly to the 6 free components of the symmetric tensor of inverse effective--mass and to the 6 symmetry elements (vertices) to be determined. Finally we make a series expansion in $\vk$, in order to get the angular dependence of the vertices on the Fermi surface.
Our results for the tetragonal point group C$_{4v}$ are
\begin{subequations} \label{App:Eq_vertex_C4v}
\begin{align}
    \gamma^{(R)}_{A_1} &= \sum\limits_{k=0}^\infty \sum\limits_{l=0}^{l\leq k/2} \gamma^{(R)}_{k,l}\cos4l\phi\, \sin^{2k}\theta \\
    \gamma^{(R)}_{B_1} &= \sum\limits_{k=1}^\infty \sum\limits_{l=1}^{l\leq (k+1)/2} \gamma^{(R)}_{k,l}\cos(4l-2)\phi\, \sin^{2k}\theta  \\
    \gamma^{(R)}_{B_2} &= \sum\limits_{k=1}^\infty \sum\limits_{l=1}^{l\leq (k+1)/2} \gamma^{(R)}_{k,l}\sin(4l-2)\phi\, \sin^{2k}\theta  \\
    \gamma^{(R)}_{E} &= \sum\limits_{k=1}^\infty \sum\limits_{l=1}^\infty \gamma^{(R)}_{k,l}\sin(2l-1)\phi\, \sin 2k\theta
\end{align}
\end{subequations}
and for the cubic point group $O$ we obtain
\begin{subequations} \label{App:Eq_vertex_O432}
\begin{align}
    \gamma^{(R)}_{A_1} &= \sum\limits_{k=0}^\infty \sum\limits_{l=0}^{l\leq k/2} \gamma^{(R)}_{k,l}\cos4l\phi\, \sin^{2k}\theta \\
    \gamma^{(R)}_{E^{(1)}} &= \gamma^{(R)}_0 (2-3\sin^2\theta) + \ldots  \\
    \gamma^{(R)}_{E^{(2)}} &= \sum\limits_{k=1}^\infty \sum\limits_{l=1}^{l\leq (k+1)/2} \gamma^{(R)}_{k,l}\cos(4l-2)\phi\, \sin^{2k}\theta  \\
    \gamma^{(R)}_{T_2} &= \sum\limits_{k=1}^\infty \sum\limits_{l=1}^{l\leq (k+1)/2} \gamma^{(R)}_{k,l}\sin(4l-2)\phi\, \sin^{2k}\theta
\end{align}
\end{subequations}
in a backscattering--geometry experiment ($z\overline{z}$).


\end{document}